\documentclass[preprint]{aastex631}

\usepackage{enumitem}
\renewcommand{\textbf}{}

\received{October 24, 2021}
\revised{January 16, 2022}
\accepted{January 23, 2022}

\shorttitle{A model $\alpha$-Cen-Earth}
\shortauthors{Wang et al.}
\graphicspath{{./}{figures/}}

\begin{document}

\title{A model Earth-sized planet in the habitable zone of $\alpha$ Centauri A/B}

\correspondingauthor{Haiyang S. Wang}
\email{haiwang@phys.ethz.ch}

\author[0000-0001-8618-3343]{Haiyang S. Wang}
\affiliation{ETH Z\"{u}rich, Institute for Particle Physics and Astrophysics, Wolfgang-Pauli-Strasse 27, CH-8093 Z\"{u}rich, Switzerland}
\affiliation{National Center for Competence in Research PlanetS (www.nccr-planets.ch)}

\author[0000-0003-2047-1558]{Charles H. Lineweaver}
\affiliation{Research School of Astronomy and Astrophysics, Australian National University, Canberra, ACT 2611, Australia}
\affiliation{Research School of Earth Sciences, Australian National University, Canberra, ACT 2601, Australia}

\author[0000-0003-3829-7412]{Sascha P. Quanz}
\affiliation{ETH Z\"{u}rich, Institute for Particle Physics and Astrophysics, Wolfgang-Pauli-Strasse 27, CH-8093 Z\"{u}rich, Switzerland}
\affiliation{National Center for Competence in Research PlanetS (www.nccr-planets.ch)}

\author[0000-0003-0000-125X]{Stephen J. Mojzsis}
\affiliation{Origins Research Institute, Research Centre for Astronomy and Earth Sciences, Konkoly Thege Mikl\'os \'ut 15-17, H- 1121 Budapest, Hungary.}
\affiliation{Department of Lithospheric Research, University of Vienna, UZA 2, Althanstra{\ss}e 14, A-1090 Vienna, Austria.}
\affiliation{Department of Geological Sciences, University of Colorado, 2200 Colorado Avenue, Boulder, CO 80309-0399, USA.}

\author[0000-0001-7617-3889]{Trevor R. Ireland}
\affiliation{School of Earth and Environmental Sciences, University of Queensland, St Lucia QLD 4072, Australia}
\affiliation{Research School of Earth Sciences, Australian National University, Canberra, ACT 2601, Australia}

\author[0000-0002-1462-1882]{Paolo A. Sossi}
\affiliation{ETH Z\"{u}rich, Institute of Geochemistry and Petrology, Sonneggstrasse 5, CH-8092 Z\"{u}rich, Switzerland.}

\author{Fabian Seidler}
\affiliation{ETH Z\"{u}rich, Institute for Particle Physics and Astrophysics, Wolfgang-Pauli-Strasse 27, CH-8093 Z\"{u}rich, Switzerland}

\author[0000-0002-8176-4816]{Thierry Morel}
\affiliation{Space sciences, Technologies and Astrophysics Research (STAR) Institute, Universit\'e de Li\`{e}ge, Quartier Agora, All\'ee du 6 Ao\^{u}t 19c, B\^{a}t. B5C, 4000 Li\`{e}ge, Belgium.}




\begin{abstract}

The bulk chemical composition and interior structure of rocky exoplanets are of fundamental importance to understanding their long-term evolution and potential habitability. Observations of the chemical compositions of the solar system rocky bodies and of other planetary systems have increasingly shown a concordant picture that the chemical composition of rocky planets reflects that of their host stars for refractory elements, whereas this expression breaks down for volatiles. This behavior is explained by devolatilization during planetary formation and early evolution. Here, we apply a devolatilization model calibrated with solar system bodies to the chemical composition of our nearest Sun-like stars -- $\alpha$ Centauri A and B -- to estimate the bulk composition of any habitable-zone rocky planet in this binary system (``$\alpha$-Cen-Earth"). Through further modeling of likely planetary interiors and early atmospheres, we find that compared to Earth, such a planet is expected to have (i) a reduced (primitive) mantle that is similarly dominated by silicates albeit enriched in carbon-bearing species (graphite/diamond); (ii) a slightly larger iron core, with a core mass fraction of $38.4_{-5.1}^{+4.7}$ wt\% (cf. Earth's 32.5 $\pm$ 0.3 wt\%); (iii) an equivalent water-storage capacity; and (iv) a CO$_2$-CH$_4$-H$_2$O-dominated early atmosphere that resembles that of Archean Earth. Further taking into account its $\sim$ 25\% lower intrinsic radiogenic heating from long-lived radionuclides, an ancient $\alpha$-Cen-Earth ($\sim$ 1.5-2.5 Gyr older than Earth) is expected to have less efficient mantle convection and planetary resurfacing, with a potentially prolonged history of stagnant-lid regimes.

\end{abstract}



\section{Introduction} \label{sec:intro}
To date, over 4900 exoplanets have been confirmed \citep{ps}\footnote{Accessed on 2022-01-16 at 18:00.} and exoplanet statistics suggest that close to 100\% of Sun-like (FGK) stars harbor planetary systems \citep{Cassan2012, Lineweaver2012, Winn2015, Zhu2021}. The occurrence of planets with radii between 0.5 and 1.5 $R_{\oplus}$ orbiting in the conservative habitable zone (HZ) \citep{Kopparapu2013} of stars with effective temperatures between 4800 and 6300 K, $\eta_{\oplus}$, is estimated to be at least $0.37_{-0.21}^{+0.48}$ \citep{Bryson2021}. Discoveries continue with legacy data from missions like Kepler, ongoing space observations by TESS, and ground-based radial velocity surveys (e.g., HARPS, CARMENES). Available data for presumably rock-dominated exoplanets, however, are typically limited to mass and/or radius, and orbital parameters. The James Webb Space Telescope (JWST), \textbf{launched} in December 2021, may be able to detect the atmospheres of a few transiting terrestrial exoplanets orbiting within the HZ of nearby M-dwarf stars \citep{Koll2019}. Towards the end of the 2020s, mid-infrared instruments installed on the 30-40 m Extremely Large Telescope (ELT) are expected to be able to directly image terrestrial, HZ exoplanets around the very nearest FGK stars \citep{Quanz2015, Bowens2021}. In anticipation of these observations and to guide the development of even more ambitious future space mission concepts, such as HabEx \citep{Gaudi2020}, LUVOIR \citep{LUVOIR2019}, and LIFE \citep{Quanz2021}, we must \textbf{focus} on modeling efforts for predictions of rocky exoplanet bulk compositions and interiors. 

Observations of the chemical compositions of rocky bodies in the Solar System \citep{Wang2019a, Sossi2018, Carlson2014, Grossman1974} and of polluted white dwarfs \citep{Harrison2021, Harrison2018, Doyle2019} lend support to the idea that the chemical composition of “terrestrial” (silicate+metal dominated) planets generally reflects that of their host stars for refractory elements, whereas this expression breaks down for volatile elements \citep{Schulze2021, Adibekyan2021}. This discrepancy can be explained by devolatilization processes \citep{Wang2019a, Sossi2019, Norris2017, Hin2017} that occurred during the formation and early evolution of the terrestrial planets of our best star-planet sample, the Solar System. Analysis shows that a variety of both stochastic and non-stochastic physical processes account for the overall devolatilization outcome. Importantly, the average effect of these processes is generally non-random \citep{Wang2018b} and follows a trend that reflects the variation of elemental abundances with elemental volatilities \citep{Fegley2020, Wang2019a, Sossi2018}. The first quantitative model of devolatilization \citep{Wang2019a} is \textbf{being} increasingly adopted in the modeling of rocky exoplanetary bulk compositions and interiors \citep{Wang2019b, Acuna2019, Spaargaren2020} or as a reference for linking star-planet elemental relations \citep{Dorn2019, Liu2020, Cowley2021, Clark2021, Schulze2021}. Here, we apply this technique to a key target for exoplanet surveys -- the $\alpha$ Centauri AB System (hereafter, $\alpha$-CenA/B) -- our nearest Sun-like stellar neighbors.

\subsection{The status of the search for an $\alpha$-CenA/B planet}
\begin{figure*}
	\includegraphics[trim=0cm 0cm 0cm 0cm, scale=0.65,angle=0]{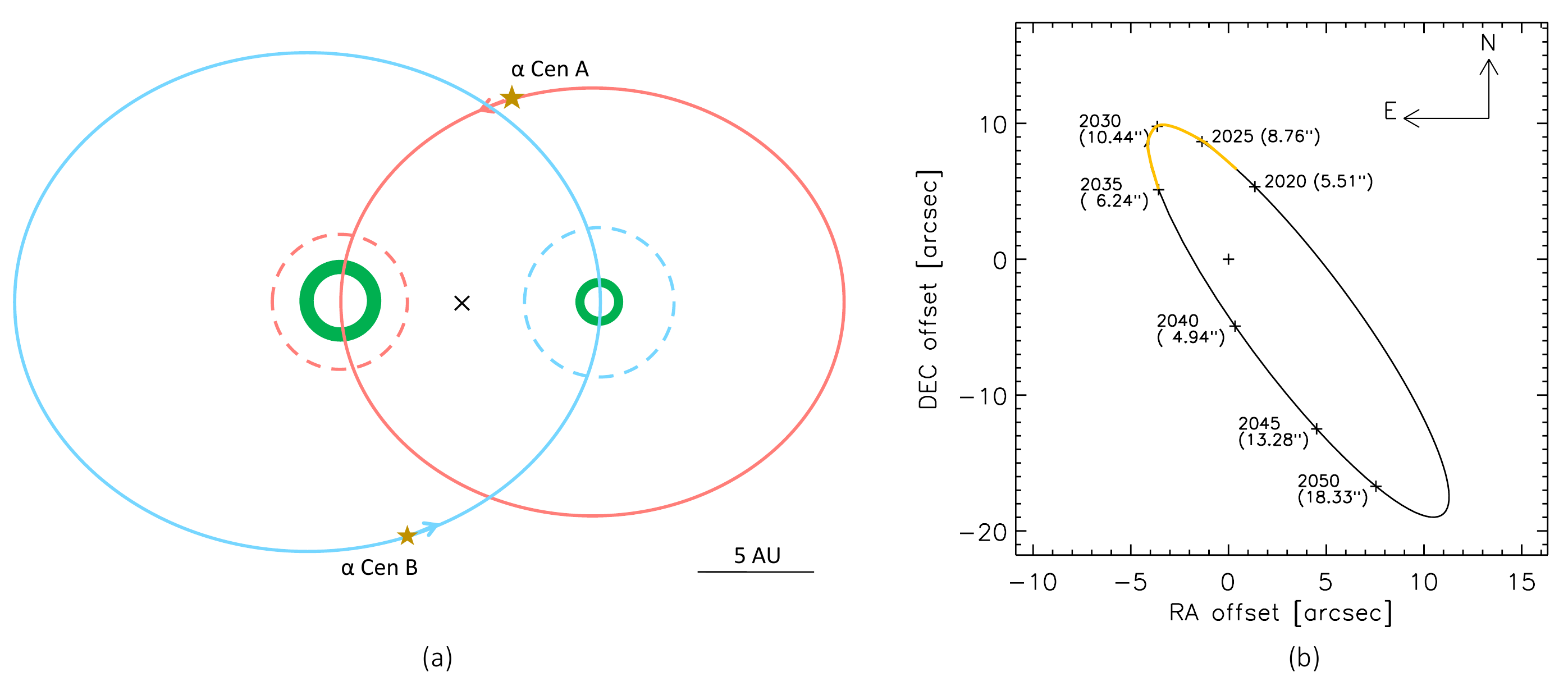} 
	\caption{(a) Trajectories of $\alpha$-Cen A (red) and B (blue) around their barycenter (``$\times$"). The two stars are positioned at their approximate present-day separation. The Hill spheres (dashed circles) and habitable zones (nested green circles) of A and B are drawn to scale at periapsis. (b) Apparent trajectory of B centered on A, with indications of their apparent separation on the sky over the period from CE 2020 to 2050. The part of trajectory in yellow indicates the coming observational window (CE 2022-2035), when the apparent separation between A and B is larger than 6$^{\prime\prime}$ and the search for planets around A or B can be conducted without suffering significant contamination from the respective companion star.} 
	\label{fig:orbits}
\end{figure*}

The binary stars of the system -- $\alpha$-Cen A (spectral type G2) and B (spectral type K1) -- are on an inclined ($\sim79^{\circ}$), eccentric ($\sim 0.5$), $\sim 80$ year orbit with a separation varying from 35.6 AU to 11.2 AU (\citealt{Pourbaix2016}; Fig. \ref{fig:orbits}) and have an estimated age of $\sim$ 6--7 Gyr \citep{Salmon2021, Morel2018}. The binary shares a gravitationally-bound companion star -- the red dwarf $\alpha$-Cen C (Proxima Centauri; spectral type M5), which orbits the barycentre of the A/B system with a period of $\sim$ 550 kyr \citep{Kervella2017}. Proxima Cen is confirmed to host at least one planet \citep{Anglada2016, Damasso2020} whereas such confirmation remains elusive for $\alpha$-CenA/B. The ostensible detection of an Earth-mass exoplanet orbiting $\alpha$-Cen B \citep{Dumusque2012} was later found to be an artifact of the observational technique \citep{Rajpaul2016}. However, owing to its immediate proximity to our Solar System, $\alpha$-CenA/B continues to generate interest in the search for Earth-like planets \citep{Zhao2018, Kasper2019, Beichman2020, Wagner2021, Akeson2021}. Radial-velocity measurements put the following detection limits on the masses of planets in the habitable zones of A and B, respectively: $M_A\sin i  < 53 M_{\oplus}$ and $M_B\sin i  < 8.4 M_{\oplus}$ \citep{Zhao2018}. Recent attempts to directly image low-mass planets in $\alpha$-CenA/B with the NEAR experiment at the Very Large Telescope (VLT) observatory have ruled out Jupiter-sized planets \citep{Wagner2021}. A candidate signal that may be consistent with an exoplanet of $R \sim 3.3 - 7 R_{\oplus}$ around $\alpha$-Cen A awaits confirmation \citep{Wagner2021}. 

Simultaneously, numerical simulations continue to support the idea that over geological timescales (i.e. $> 10^9$ yr) stable planetary orbits can persist within the habitable zones of both $\alpha$-Cen A and 
B \citep{Andrade2014, Quarles2016, Quarles2020}. Such stability is also bolstered by the non-overlapping Hill spheres of the binary stars at their periapsis (Fig. \ref{fig:orbits}a). The interval between the years 2022--2035 will be the next ideal observational timeframe (Fig. \ref{fig:orbits}b), during which searching for planets around A or B with, e.g., JWST/MIRI \citep{Beichman2020} and ELT/METIS \citep{Quanz2015, Carlomagno2020}, will benefit from lower levels of contamination from the other component. It is therefore timely to predict what kind of planets to expect in the $\alpha$-CenA/B system based on theoretical models, and thus provide guidance for future observations. For this study, we focus our analysis on a hypothetical, Earth-sized planet in the habitable zone around either A or B.

\section{Data, methods, and analysis} \label{sec:methods}
\subsection{Stellar chemical compositions}\label{sec:data}
\begin{figure*}
	\centering
	\includegraphics[trim=0cm 0cm 0cm 0cm, scale=0.65,angle=0]{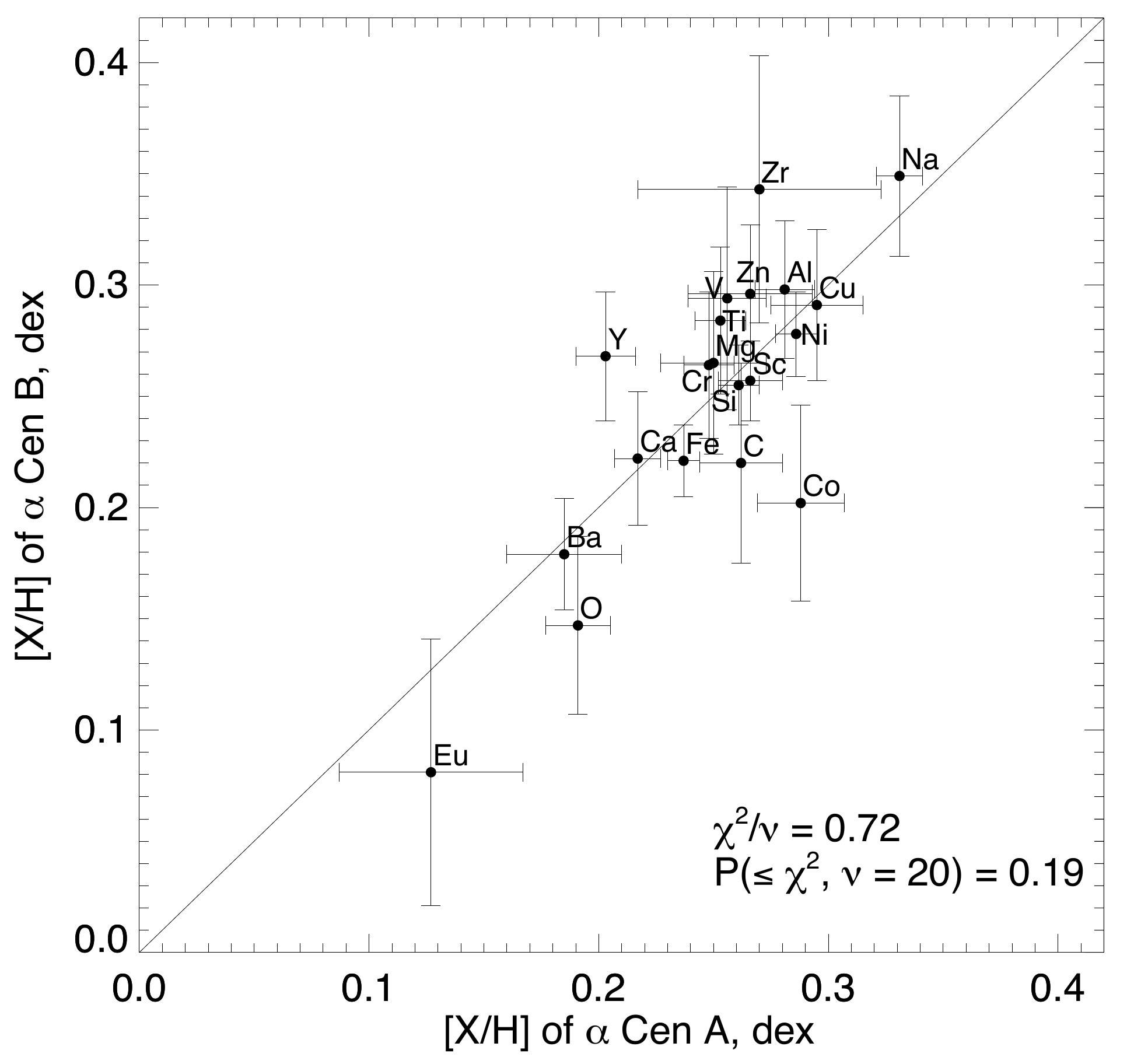} 
	\caption{Comparison of the elemental abundances of $\alpha$-Cen A and B. A $\chi^2$-test shows that $\chi^2/\mathbf{\nu} = 0.72$ and $P(\ge\chi^2, \mathbf{\nu}) = 0.19$, where the degrees of freedom $\mathbf{\nu} = 20$ (i.e., the number of elements whose abundances are available in both A and B). The data sources are from \cite{Morel2018} and \cite{Wang2020} and are also listed in Table \ref{tab:starabu}.} 
	\label{fig:ABcomp}
\end{figure*}

Both $\alpha$-Cen A and B are among the Gaia FGK “benchmark stars”, for which the stellar properties have been carefully calibrated \citep{Heiter2015,Jofre2017}. The chemical compositions of the two companions have also been determined in detail with high-quality spectra for up to 22 elements \citep{Morel2018, Wang2020}. The datasets of \citep{Morel2018, Wang2020} are selected for (i) their completeness, high precision and homogeneity for all elements of interest to our study and (ii) their large consistency with other reported results in the literature for $\alpha$ Cen A/B \citep[e.g.][]{Neuforge1997, Koch2002, Guiglion2018}. We do not compare the different literature analyses here, and such comparisons have been discussed in detail in our refereed sources.  

Figure \ref{fig:ABcomp} compares the elemental abundances between $\alpha$-Cen A and B (Columns 2-3, Table \ref{tab:starabu}), which are found to be statistically consistent with each other ($\chi^2/\nu= 0.72$ and $P(\le\chi^2,\nu)=0.19$), supporting the idea that A and B have a common origin \citep{Beech2015}. For consistency and simplicity of discussion, we calculate the weighted-average elemental abundances of A and B (Column 4, Table \ref{tab:starabu}). 
The exceptions to the approach are that (i) since the A and B abundances of Co and Y are inconsistent, we compute unweighted averages and use their abundance spread as uncertainties; (ii) the abundances of Mn and Ce are only available for $\alpha$ Cen A and are thus adopted directly for the averaged AB system. We then use the protosolar abundances \citep{Wang2019a} to convert the differential abundances to the absolute abundances (normalized to Al = 100; Column 5, Table \ref{tab:starabu}). We did not consider the effects of microscopic diffusion in $\alpha$ Cen A and B, but as noted in \cite{Morel2018} any changes arising from diffusion can probably be accommodated by the abundance uncertainties. 

\begin{table}
	\caption{\label{tab:starabu} The elemental abundances of $\alpha$ Cen A, $\alpha$ Cen B, and the average AB system. For details see Sect. \ref{sec:data}.}
	\begin{tabular}{clllr@{}l}	
	\hline\hline
		Element & [X/H]$_\textrm{A}$ (dex) & [X/H]$_\textrm{B}$ (dex) & [X/H]$_{\textrm{AB}}$ (dex) & \multicolumn{2}{c}{(X/Al)$_{\textrm{AB}}$ (Al=100)}\\
		\hline
		C &	$0.262 \pm 0.018$	& $0.220 \pm 0.045$	& $0.256 \pm 0.017$	& 9514&$\,_{-1106}^{+1251}$ \\
		O&	0.191 $\pm$ 0.014&	0.147 $\pm$ 0.040&	0.186 $\pm$ 0.013&	14735&$\,_{-1683}^{+1900}$\\
		Na&	0.331     $\pm$ 0.010&	0.339 $\pm$ 0.036&	0.332 $\pm$ 0.010&	72&$.7\,_{-4.8}^{+5.1}$\\
		Mg&	0.250 $\pm$ 0.023&	0.265 $\pm$ 0.041&	0.254 $\pm$ 0.020&	1192&$\,_{-85}^{+318}$\\
		Al&	0.281  $\pm$ 0.013&	0.298 $\pm$ 0.031&	0.284 $\pm$ 0.012&	100 & $\,\pm 6$\\
		Si&	0.261 $\pm$ 0.009&	0.255 $\pm$ 0.018&	0.260 $\pm$ 0.008&	1153&$\,_{-38}^{+39}$\\
		Ca&	0.217 $\pm$ 0.010&	0.222 $\pm$ 0.030&	0.218 $\pm$ 0.009&	64&$.2\,_{-3.2}^{+3.4}$\\
		Sc&	0.266 $\pm$ 0.014&	0.257 $\pm$ 0.018&	0.263 $\pm$ 0.011&	0&$.042\,_{-0.005}^{+0.016}$\\
		Ti&	0.253 $\pm$ 0.011&	0.284 $\pm$ 0.033&	0.256 $\pm$ 0.010&	2&$.86\,_{-0.17}^{+0.18}$\\
		V&	0.256 $\pm$ 0.017&	0.294 $\pm$ 0.050&	0.260 $\pm$ 0.016&	0&$.32 \,\pm 0.02$\\
		Cr&	0.248 $\pm$ 0.011&	0.264 $\pm$ 0.033&	0.250 $\pm$ 0.010&	14&$.8_{-0.7}^{+0.8}$\\
		Mn&	0.271 $\pm$ 0.019&	--&	0.271 $\pm$ 0.019&	10&$.5 \,\pm 0.7$\\
		Fe&	0.237 $\pm$ 0.007&	0.221 $\pm$ 0.016&	0.234 $\pm$ 0.006&	959&$\,_{-37}^{+38}$\\
		Co&	0.288 $\pm$ 0.019&	0.202 $\pm$ 0.044&	$0.245\;_{-0.087}^{+0.062}$&	2&$.57\,_{-0.47}^{+0.41}$\\
		Ni&	0.286 $\pm$ 0.009&	0.278 $\pm$ 0.019&	0.285 $\pm$ 0.008&	59&$.6\,_{-3.4}^{+3.6}$\\
		Cu&	0.295 $\pm$ 0.020&	0.291 $\pm$ 0.034&	0.294 $\pm$ 0.017&	0&$.62 \,\pm 0.06$\\
		Zn&	0.266 $\pm$ 0.027&	0.296 $\pm$ 0.031&	0.279 $\pm$ 0.020&	1&$.48 \,\pm 0.09$\\
		Y&	0.203 $\pm$ 0.013&	0.268 $\pm$ 0.029&	$0.236\;_{-0.046}^{+0.062}$&	4&$.78\,_{-0.52}^{+0.77} \;\times 10^{-3}$\\
		Zr&	0.270 $\pm$ 0.053&	0.343 $\pm$ 0.060&	0.302 $\pm$ 0.040&	0&$.014 \pm 0.001$\\
		Ba&	0.185 $\pm$ 0.025&	0.179 $\pm$ 0.025&	0.182 $\pm$ 0.018&	4&$.50\,_{-0.27}^{+0.29} \;\times10^{-3}$\\
		Ce&	0.172 $\pm$ 0.055&	--	&0.172 $\pm$ 0.055&	1&$.09\,_{-0.13}^{+0.15} \;\times10^{-3}$\\
		Eu&	0.127 $\pm$ 0.040&	0.081 $\pm$ 0.060&	0.113 $\pm$ 0.033&	8&$.36\,_{-0.66}^{+0.71} \;\times10^{-5}$\\
		
		\hline
	\end{tabular}
\end{table}

\subsection{Devolatilization}\label{sec:devol}
We employ the fiducial model of devolatilization \citep{Wang2019a}, which is quantified by the bulk elemental abundance differences between the proto-Sun and the Earth \citep{Wang2018, Wang2019a} as a function of 50\% condensation temperature \citep[$T_c$$^{50}$;][]{Lodders2003}, to devolatilize the averaged AB system abundances. This process results in the devolatilized host stellar abundances (Column 2, \textbf{Table} \ref{tab:planetabu}) as the proxy for the elemental composition of any rocky planet in the HZ of the system (i.e. “$\alpha$-Cen-Earth”). The justification and limitations of applying such a Sun-to-Earth model to other Solar-like planetary systems are presented in \cite{Wang2019b} and also recapped later. Further, since we are interested in how the model $\alpha$-Cen-Earth is different from the bulk Earth by analyzing primarily the individual elements or their ratios, the residuals of the devolatilization model for individual elements (Column 3, Table \ref{tab:planetabu}) -- i.e. the differences between the bulk Earth composition \citep{Wang2018} and the model Earth composition as devolatilized from the protosolar abundances \citep{Wang2019a} -- are added onto the devolatilized AB average stellar abundances to obtain the final bulk elemental composition of the model $\alpha$-Cen-Earth (Columns 4-5, Table \ref{tab:planetabu}). 

\begin{table}
	\caption{\label{tab:planetabu} Estimates of the chemical composition of the model $\alpha$-Cen-Earth.  For details see Sect. \ref{sec:devol}.}
	\centering
	\begin{tabular}{c r@{}l r@{}l r@{}l r@{}l}
		\hline\hline
		Element&	\multicolumn{2}{c}{(X/Al)$_{\textrm{devol}}$ (Al=100)}&	\multicolumn{2}{c}{(X/Al)$_{\textrm{residual}}$}&	\multicolumn{2}{c}{(X/Al)$_{\textrm{bulk}}$ (Al=100)}&	\multicolumn{2}{c}{(X/Fe)$_{\textrm{bulk}}$ (Fe=100)}\\
		\hline
		C&	35&$.9\,_{-7.7}^{+9.0}$&	0&.3& 36&$.2\,_{-7.7}^{+9.0}$&	4&$.12\,_{-0.87}^{+1.03}$\\
		O&	2674&$\,_{-335}^{+377}$&	13&&	2687&$\,_{-335}^{+377}$&	306&$\,_{-38}^{+43}$\\
		Na&	18&$.4\,_{-1.4}^{+1.6}$&	0&.3&	18&$.7\,_{-1.4}^{+1.6}$&	2&$.13\,_{-0.16}^{+0.18}$\\
		Mg&	1025&$\,_{-81}^{+276}$&	-15&&	1010&$\,_{-81}^{+276}$&	115&$\,_{-9}^{+31}$\\
		Al&	100&$\,\pm 6$&	0&&	100&$\;\pm 6$&	11&$.4\,_{-0.6}^{+0.7}$\\
		Si&	923&$\,_{-43}^{+47}$&	19&&	941&$\,_{-43}^{+47}$&	107&$\,\pm 5$\\
		Ca&	64&$.2\,_{-3.2}^{+3.4}$&	-1&.8&	62&$.4\,_{-3.2}^{+3.4}$&	7&$.11\,_{-0.36}^{+0.38}$\\
		Sc&	0&$.042\,_{-0.005}^{+0.016}$&	-0&.003&	0&$.039\,_{-0.004}^{+0.015}$&	0&$.004\,_{-0.001}^{+0.002}$\\
		Ti&	2&$.9 \,\pm 0.2$&	-0&.2&	2&$.7\,\pm  0.2$&	0&$.31\,\pm 0.02$\\
		V&	0&$.32\,\pm 0.02$&	-0&.01&	0&$.31\,\pm  0.02$&	0&$.036\,_{-0.002}^{+0.003}$\\
		Cr&	11&$.4\,\pm  0.7$&	0&.7&	12&$.1\,\pm 0.7$&	1&$.38\,\pm  0.08$\\
		Mn&	5&$.3\,\pm  0.4$&	0&.1&	5&$.4\,\pm  0.4$&	0&$.62\,_{-0.04}^{+0.05}$\\
		Fe&	820&$\,_{-42}^{+46}$&	57&&	877&$\,_{-42}^{+46}$&	100&$\,\pm  5$\\
		Co&	2&$.3\,\pm  0.4$&	0&.0&	2&$.3\,\pm  0.4$&	0&$.26_{-0.05}^{+0.04}$\\
		Ni&	53&$.7\,_{-3.6}^{+3.9}$&	-0&.8&	52&$.9_{-3.6}^{+3.9}$&	6&$.03\,_{-0.41}^{+0.44}$\\
		Cu&	0&$.21\, \pm 0.02$&	-0&.02&	0&$.19\,\pm  0.02$&	0&$.022\,_{-0.002}^{+0.003}$\\
		Zn&	0&$.14\, \pm 0.01$&	-0&.01&	0&$.12\,\pm  0.01$&	0&$.014\,_{-0.001}^{+0.002}$\\
		Y&	4&$.77\,_{-0.52}^{+0.77} \;\times10^{-3}$&	0&$.03\;\times10^{-3}$&	4&$.81\,_{-0.52}^{+0.77} \;\times{10}^{-3}$&	5&$.48\,_{-0.59}^{+0.87}\;\times10^{-4}$\\
		Zr&	0&$.014\,\pm  0.001$&	0&.000&	0&$.014\,\pm  0.001$&	1&$.54\,_{-0.15}^{+0.17}\;\times 10^{-3}$\\
		Ba&	4&$.50\,\pm_{-0.27}^{+0.29} \;\times10^{-3}$&	-0&$.34\;\times10^{-3}$&	4&$.16\,_{-0.25}^{+0.27} \;\times10^{-3}$& 4&$.75\,_{-0.31}^{+0.33} \;\times10^{-4}$\\
		Ce&	1&$.09\,_{-0.13}^{+0.15} \;\times10^{-3}$&	-0&$.03\;\times10^{-3}$&	1&$.06\,_{-0.13}^{+0.15} \;\times10^{-3}$&	1&$.21\,_{-0.15}^{+0.17}\;\times10^{-4}$\\
		Eu&	7&$.59\,_{-0.65}^{+0.71} \;\times10^{-5}$&	0&$.41 \;\times10^{-5}$&	8&$.01\,_{-0.65}^{+0.71} \;\times10^{-5}$&	9&$.1_{-0.7}^{+0.8}\;\times10^{-6}$\\
		\hline
	\end{tabular}
\end{table}

\subsection{Key geochemical ratios}\label{sec:ratios}
The stellar abundance ratios of carbon to oxygen (C/O) and magnesium to silicon (Mg/Si) are diagnostic indicators, as widely adopted in the literature \citep[e.g.][]{Bond2010, Delgado2010, Fortney2012, Thiabaud2015, Brewer2016, Suarez-Andres2018, Zhao2018, Spaargaren2020, Clark2021} (we also use them later), of the plausible composition of any rocky planet around the host star. However, upon the application of our devolatilization approach, both carbon and oxygen are severely depleted in rocky planets relative to their host star and therefore, a planetary (not a stellar) C/O is no longer a valid indicator of rocky planets' dominant mineral types (silicates vs. carbides). Due to the \textbf{near-}refractory nature (and similar $T_c$$^{50}$; \citealt{Lodders2003, Wood2019}) of Mg and Si, their ratio in a star is not significantly altered through the devolatilization process \citep{Wang2019a} and thus is still a good first-order proxy for Mg/Si in a rocky planet around the host star \citep{Schulze2021, Adibekyan2021, Thiabaud2014, Delgado2010, Bond2010}.   

The oxidation state of a planet is crucial to understand planetary chemistry and internal fractionation of materials \citep{Wade2005, Unterborn2014} and it is usually recorded by oxygen fugacity ($f\textrm{O}_2$) \textbf{-- i.e. the nonideal partial pressure of oxygen. It may, for convenience, be expressed relative to a mineral buffer, such as Iron-W\"ustite (IW) and Quartz-Fayalite-Magnetite (QFM) \citep{Oneill1987, Ballhaus1990, Doyle2019}.} 
\textbf{In an exoplanetary context, however, the $f\textrm{O}_2$ is difficult to estimate, because it changes in response to mineralogy, pressure and temperature \citep{Woodland2003, ONeill2006, Armstrong2019}, all of which are unknown in the first place for our model $\alpha$-Cen-Earth.} 
We therefore opt for a proxy -- that may be constrained directly from the estimated planetary bulk composition -- for the planetary nominal $f\textrm{O}_2$. Considering that O, Mg, Si, and Fe are the most abundant elements in a silicate planet \citep{Wang2018, Yoshizaki2020} and that MgO and SiO$_2$ are the major oxides while FeO content heavily depends on the planet's oxidation state, we propose the bulk (O-Mg-2Si)/Fe as a proxy for the oxidation state of such a planet. For example, if a negative value is found for (O-Mg-2Si)/Fe (indeed for our case as shown in Table \ref{tab:ratios}), it implicitly reveals a reduced nature of the mantle.   

Furthermore, since Mg and Si are expected to be the dominant lithophile elements in a silicate mantle \citep{Palme2014b, Wang2018, Yoshizaki2020} while Fe is the major constituent for a rocky planet's core \citep{McDonough2003, Wang2018}, (Mg+Si)/Fe is adopted as a proxy for the potential core size of a rocky planet. Along these lines, since Eu is also a lithophile element and by following \citep{Wang2020} it is a preferred proxy for long-lived radionuclides $^{235,238}$U and $^{232}$Th in terrestrial-type planets, we therefore adopt Eu/(Mg+Si) as a proxy for indicating the concentration of these long-lived isotopes and the budget of the resultant radiogenic heating production over geologic timescales. A similar approach awaits formulation, however, to estimate $^{40}$K abundance \citep{Gastis2020}. Following \cite{Frank2014}, we assume that the $\alpha$-CenA/B system has an initial value of $^{40}$K as that of the Solar System, but we also note that due to the age of the stars in the $\alpha$-CenA/B system \citep{Morel2018, Salmon2021}, $^{40}$K is considerably depleted in the modern mantle of our model $\alpha$-Cen-Earth \citep{Wang2020}.  

\subsection{Software package and implementation}
To carry out the analysis, three sets of software are employed: \textit{ExoInt} \citep{Wang2019b} for both devolatilization and first-principal interior modeling; \textit{Perple$\_$X} \citep{Connolly2009} for a detailed modeling of the mantle mineralogy and interior structure; and \textit{FactSage 8.0} \citep{Bale2016} for calculating the gas speciation from the modeled interiors. The details of the software package and its implementation for the study are presented in Appendix \ref{sec:software}. 

\section{Results}
\subsection{The chemical composition of the average AB system}\label{sec:ABave}

\begin{figure*}
	\centering
	\includegraphics[trim=0cm 0cm 0cm 0cm, scale=0.65,angle=0]{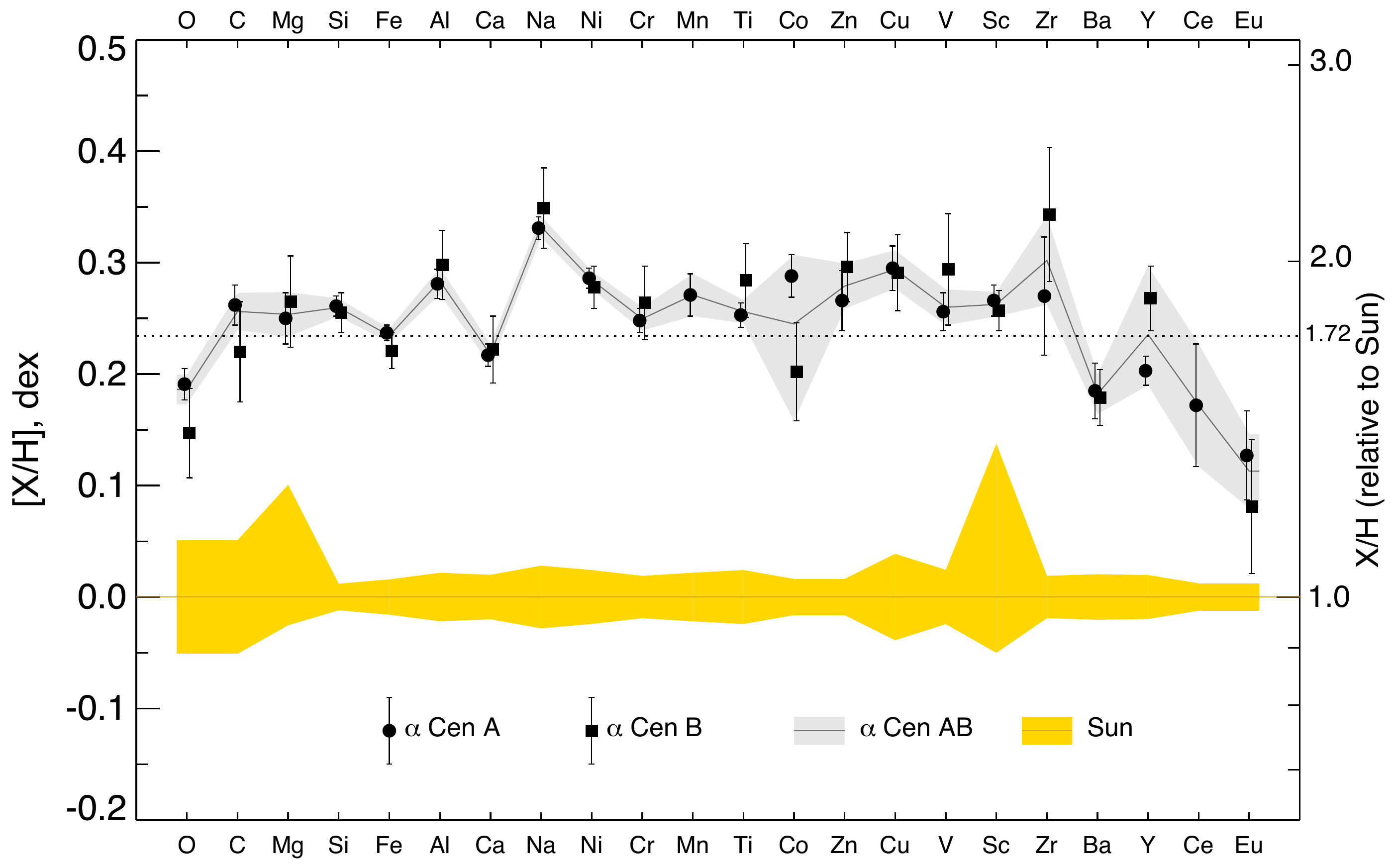} 
	\caption{Elemental abundances ([X/H], dex) of $\alpha$-Cen A and B and of the averaged AB system, in comparison with the protosolar abundances \citep{Wang2019a}. The dotted line indicates the mean enrichment ($\sim 1.72$) in metallicity (Fe/H) in the averaged AB system compared to the Sun. Elements are plotted from left to right in order of decreasing solar abundance. For more details see Sects. \ref{sec:data} and \ref{sec:ABave}.} 
	\label{fig:ABave}
\end{figure*}

Figure \ref{fig:ABave} compares the chemical composition of the average AB system with that of the Sun. First of all, the average AB system is super-solar in metallicity ([Fe/H]), with an enrichment of iron to hydrogen by $\sim$ 72\%. The enrichment of other elements, except O, Ba, Ce, and Eu, to hydrogen ([X/H]) are either equivalent or higher. The variation in the hydrogen-normalised enrichments for major rock-forming elements (e.g., O, Mg, Si, Fe, Al, Ca, and Na) will have a profound influence, as demonstrated later, on the geochemistry of the model $\alpha$-Cen-Earth.

\begin{figure}
	\centering
	\includegraphics[trim=0cm 0cm 0cm 0cm, scale=0.5,angle=0]{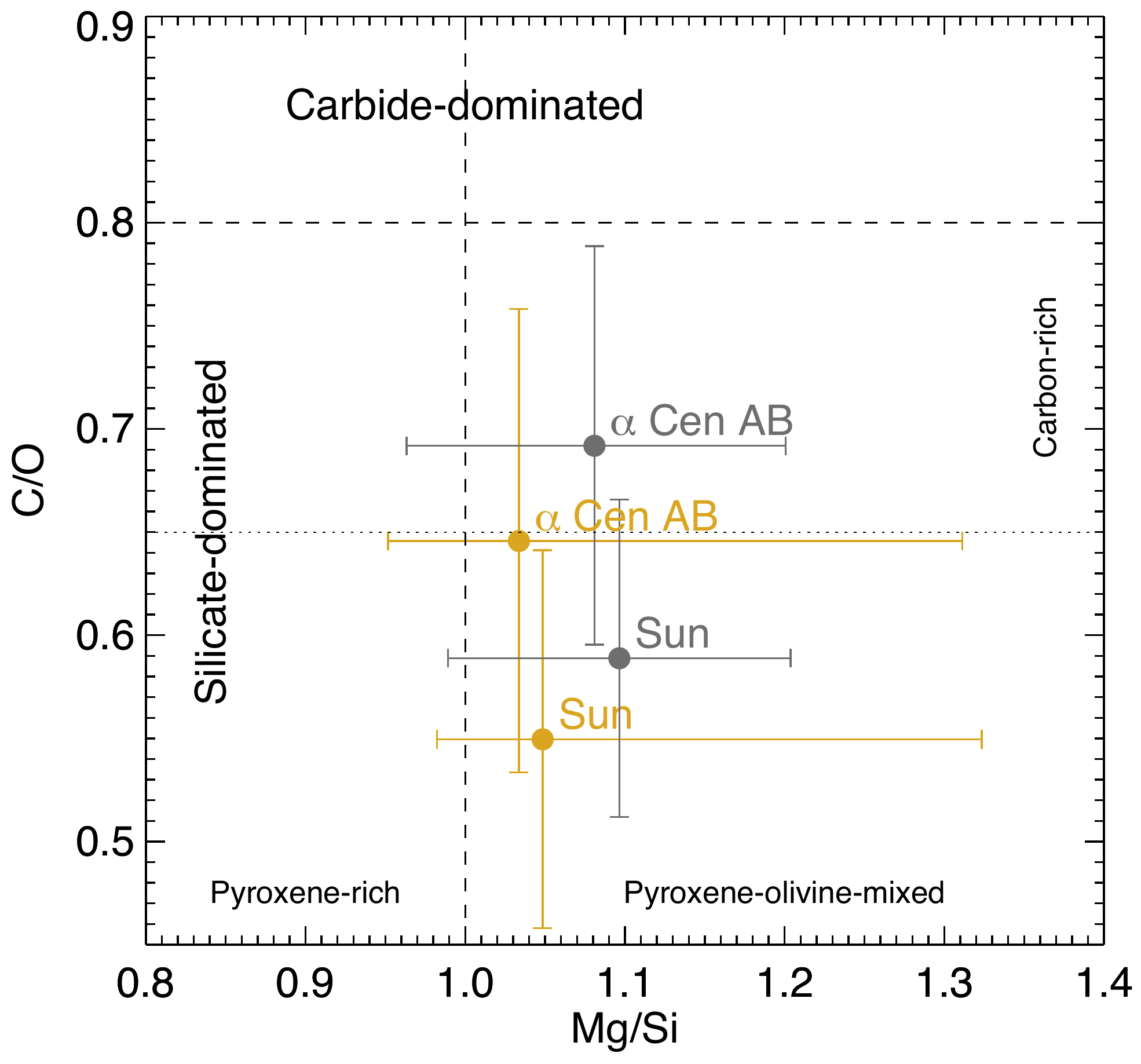} 
	\caption{The comparison of $\alpha$ Cen AB with the Sun on the C/O--Mg/Si diagram, with two different sets of solar abundances for references: dots in yellow -- \cite{Wang2019a}; dots in grey -- \cite{Asplund2021}. The dashed vertical and horizontal lines indicate the canonical boundaries for the mineralogy of HZ rocky planets in a Sun-like star system \citep{Bond2010, Delgado2010}, while the dotted horizontal line indicates a lower limit of C/O, above which carbon-bearing species is enriched in a silicate-dominated mantle \citep{Brewer2016}.} 
	\label{fig:star_ratio}
\end{figure}

To first order, C/O affects the availability of rock-forming elements such as Mg and Si to bond with O (e.g. Mg$_2$SiO$_4$) or C (e.g. SiC), while Mg/Si modulates the silicate mineral assemblages (e.g. olivine vs. pyroxene; Mg$_2$SiO$_4$ vs. MgSiO$_3$). A calculation of C/O and Mg/Si \textbf{in $\alpha$-CenA/B} (Fig. \ref{fig:star_ratio}) reveals that both of these ratios are consistent with the solar values (within uncertainties), although the C/O ($0.65\pm0.11$) of $\alpha$-CenA/B is closer -- than that ($0.55\pm0.09$) of the Sun \citep{Wang2019a} -- to the threshold (0.8) \citep{Bond2010, Delgado2010} for forming a carbide (rather than silicate) planet. With such a relatively higher value in C/O, the $\alpha$-CenA/B planet is expected to develop a silicate mantle just like terrestrial planets in our system, whereas the carbon-bearing phases (e.g. graphite and diamond) in its exo-mantle may be slightly more abundant -- cf. \citep{Zhao2018}. An approximately equal Mg/Si between $\alpha$-CenA/B and the Sun ($1.03^{+0.28}_{-0.08}$ vs. $1.04^{+0.28}_{-0.07}$) suggests a mineralogy similar to that of the Earth, in accord with earlier studies \citep[e.g.][]{Zhao2018, Morel2018}. We also note that the most recent solar photospheric abundances \citep{Asplund2021} are consistent with our adopted protosolar abundances \citep{Wang2019a} within uncertainties in terms of C/O vs. Mg/Si (Fig. \ref{fig:star_ratio}), while a normalization to the former would lead to a relatively higher C/O that favors our suggestion of the enrichment of carbon-bearing phases in the silicates-dominated mantle of the model planet.   

\subsection{The chemical composition of the model $\alpha$-Cen-Earth}\label{sec:aEarth}
\begin{figure*}
	\centering
	\includegraphics[trim=0cm 0cm 0cm 0cm, scale=0.65,angle=0]{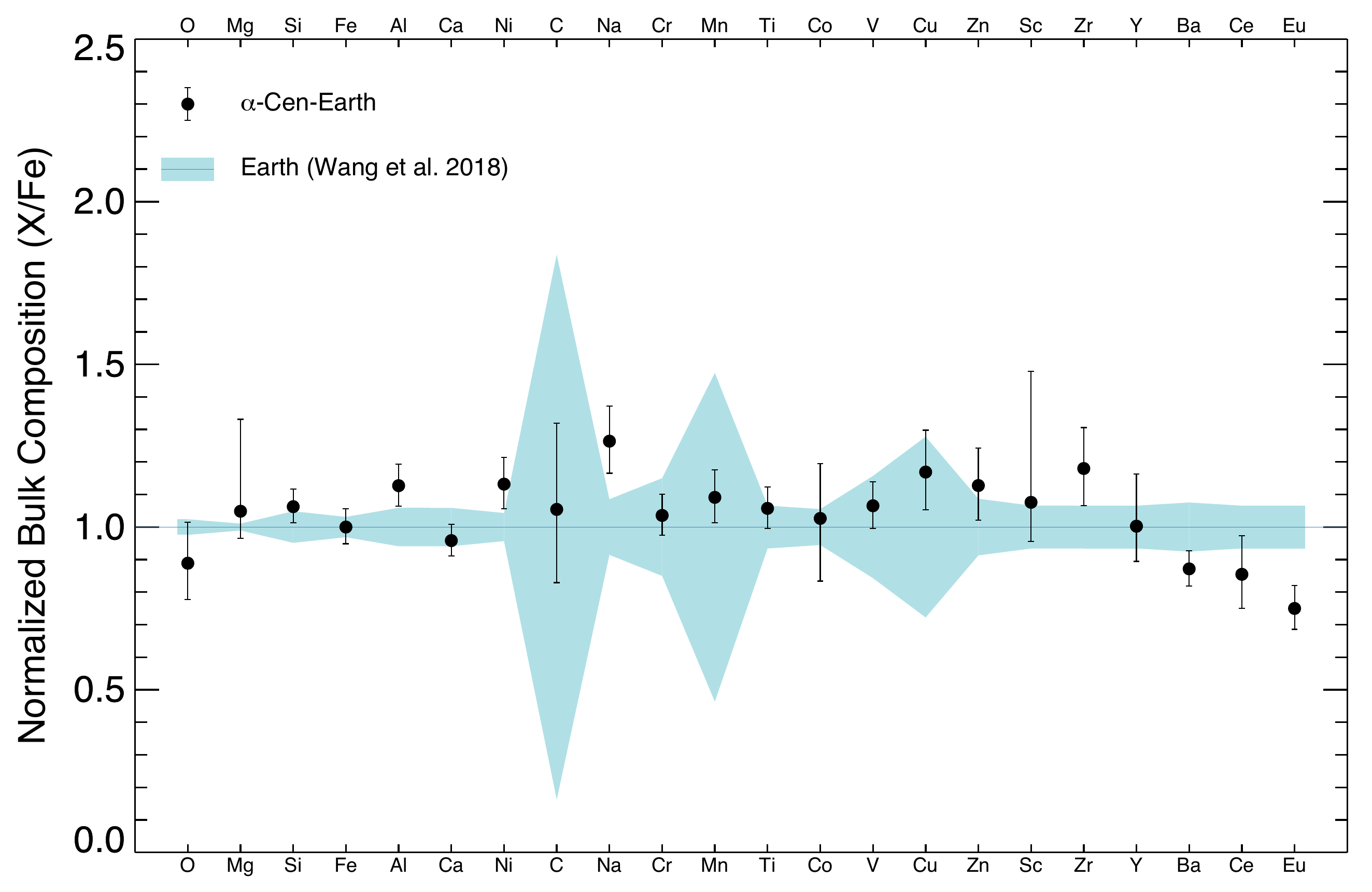} 
	\caption{Estimates of bulk elemental composition of the model $\alpha$-Cen-Earth, as devolatilized from the chemical composition of the average AB system. The elemental compositions (by number) are normalized to Fe and bulk Earth \citep{Wang2018}. The particularly large uncertainties for C, Mn, and Cu in the bulk Earth composition are caused by their large uncertainties in core composition \citep{Wang2018}. Elements are ordered by decreasing \textbf{abundance in the} bulk Earth composition. For more details see Sects. \ref{sec:devol} and \ref{sec:aEarth}.} 
	\label{fig:aEarth}
\end{figure*}
Applying the fiducial model of devolatilization \citep{Wang2019a} to the averaged elemental abundances of the $\alpha$-CenA/B system provides an estimate of the bulk elemental composition of the model $\alpha$-Cen-Earth (Table \ref{tab:planetabu}). To focus the discussion on the relative abundances of individual elements in the model $\alpha$-Cen-Earth compared to Earth, all elemental abundances are normalized to iron, and by doing so, the $\sim$ 72\% relative enrichment of metallicity is removed (Fig. \ref{fig:aEarth}). Similar to the findings shown in Fig. \ref{fig:ABave}, the model $\alpha$-Cen-Earth has equivalent or relatively higher concentrations of most elements (relative to Fe) and lower concentrations of O, Ba, Ce, and Eu. The implications of such a bulk composition for the interior chemistry, structure and dynamics of the planet are firstly analyzed with the aforementioned key geochemical ratios, the values of which for the model planet (relative to Earth) are listed in Table \ref{tab:ratios}.

\begin{table}
	\caption{\label{tab:ratios} Estimates of key geochemical ratios of the model $\alpha$-Cen-Earth (normalized to Earth, by number; \citealt{Wang2018})$^{a}$.}
	\centering
\begin{tabular}{cccc cccc}
		\hline\hline
[Mg/Si] && [(Mg+Si)/Fe] && [(O-Mg-2Si)/Fe] && [Eu/(Mg+Si)]& \\
$0.98\,_{-0.11}^{+0.28}$ && $1.04\,_{-0.08}^{+0.16}$ && $-0.79\,_{-1.74}^{+1.56}$ && $0.73\,_{-0.13}^{+0.09}$& \\
		\hline
\multicolumn{8}{p{10cm}}{$^a$ [X/Y] = (X/Y)$_{\alpha\textrm{-Cen-Earth}}$ / (X/Y)$_{\textrm{Earth}}$, where X and Y respectively denotes the numerator and denominator in any of the listed key ratios in brackets.}
\end{tabular}
\end{table}

\begin{figure*}
\centering
\includegraphics[trim=0cm 0cm 0cm 0cm, scale=0.65,angle=0]{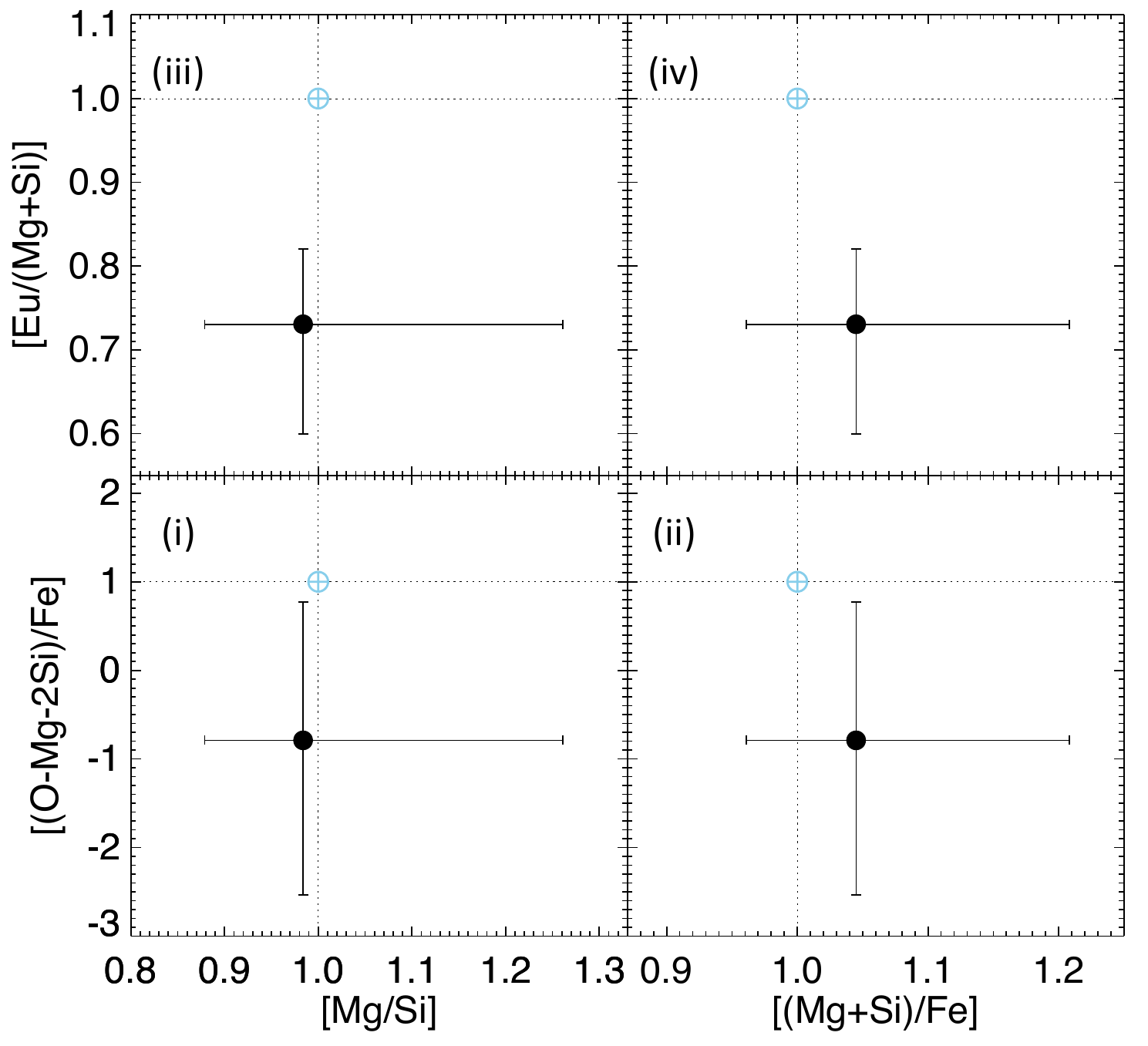} 
\caption{Key geochemical ratios of the model $\alpha$-Cen-Earth (black dots) normalized to the Earth ($\oplus$, with normalized values indicated by dotted lines; \citealt{Wang2018}). Earth's uncertainties for these geochemical ratios have been propagated to the error bars shown for the model $\alpha$-Cen-Earth. Values are also reported in Table \ref{tab:ratios}. For more details see Sect. \ref{sec:aEarth} and Appendix \ref{sec:ratios}.} 
\label{fig:planet_ratio}
\end{figure*}

Overall, the model $\alpha$-Cen-Earth's [Mg/Si] of $0.98^{+0.28}_{-0.11}$ and [(Mg+Si)/Fe] of $1.04^{+0.16}_{-0.08}$
resemble Earth's values, whereas a [(O-Mg-2Si)/Fe] of $-0.79^{+1.56}_{-1.74}$ and a [Eu/(Mg+Si)] of $0.73^{+0.09}_{-0.13}$ are relatively lower than those of Earth. Note that the uncertainties of these relative ratios include the uncertainties of these ratios in Earth. A cross-check of these key geochemical ratios, as shown in \mbox{Fig. \ref{fig:planet_ratio}}, helps understand better the geological properties of the model $\alpha$-Cen-Earth than with individual ratios. For example, in panel (i) of Fig. \ref{fig:planet_ratio}, the approximately equal [Mg/Si] reveals a silicate mantle composed of the relatively similar fractions of olivine (main formula: (Mg, Fe)$_2$SiO$_4$) and pyroxene (main formula: (Mg, Fe)SiO$_3$), which in turn may be used to argue that they have similar mantle viscosity and thus convection rate \citep{Spaargaren2020}. 
In the [(Mg+Si)/Fe] vs. [(O-Mg-2Si)/Fe] panel, a slightly higher [(Mg+Si)/Fe] alludes to a sub-equal or smaller core for the model $\alpha$-Cen-Earth, whereas a lower [(O-Mg-2Si)/Fe] points to an FeO/Fe ratio that is lower, thus resulting in an increased core size (the relative importance of these competing effects are vetted by the subsequent quantitative interior modeling). In panel (iii), a significantly lower [Eu/(Mg+Si)] -- a proxy for the concentration of the long-lived radionuclides -- that occurs in a mantle with mineralogy similar to the Earth's (suggested by [Mg/Si] above) implies that the mantle convection of the model $\alpha$-Cen-Earth may be weakened by $\sim$ 1/4 less intrinsic radiogenic heating over its geological history (assuming its gravitational energy to be similar to that of Earth; to be discussed further). Finally, a combination of panels (iv) and (ii) implies that the heat extraction from the Earth-sized core may be more efficient due to the potentially increased core-mantle temperature differences caused by the lower intrinsic radiogenic heating in the mantle \citep{Wang2020, Nimmo2020}, but also subject to the tectonic regime of the planet (to be discussed later).  

\subsection{The interior composition and structure of the model $\alpha$-Cen-Earth}\label{sec:interior}
Table \ref{tab:comp} shows the results of the modeling of first-order mantle and core composition as well as core mass fraction with \textit{ExoInt} \citep[][also see Appendix \ref{sec:exoint}]{Wang2019b}. It is found that the model $\alpha$-Cen-Earth has a mantle composition similar to that of Earth, except for its FeO content \textbf{that} is significantly depleted and native carbon species (graphite/diamond) \textbf{that may be} relatively enriched. Some portion of carbon may be fractionated into the core, but our model is unable to constrain that yet (to be discussed later).
\textbf{The results are} nonetheless consistent with the qualitative analysis above, respectively with [(O-Mg-2Si)/Fe] in the planet and with C/O in its host star. A further calculation of the Mg number (Mg\# = MgO / (MgO + FeO), in molar ratio; converted from the weight fractions in Table \ref{tab:comp}) -- an indicator for mantle chemistry and the degree of melting -- for the (primitive) mantle of the model planet reveals a much higher Mg\#, $0.987^{+0.010}_{-0.020}$, compared to $0.890 \pm 0.001$ of the Earth's primitive mantle \citep{Palme2014b, Wang2018}).  

\begin{table}
	\caption{\label{tab:comp} Estimates of mantle and core compositions as well as core mass fraction of the model $\alpha$-Cen-Earth, in comparison with those of Earth.}
	\centering
	\begin{tabular}{c cccc cccc}
		\hline\hline
{} & \multicolumn{8}{c}{Mantle composition (wt\%)}\\
\cline{2-9}
	{}	& MgO	& SiO$_2$	&FeO & Al$_2$O$_3$ &CaO&	Na$_2$O	& NiO	& C\\
$\alpha$-Cen-Earth & $44.8\,_{-7.0}^{+10.4}$	& $43.9\,_{-10.0}^{+6.4}$&	$1.0\,_{-0.8}^{+1.2}$&	$5.4 \pm 0.7$&	$3.7 \pm 0.5$	& $0.6 \pm 0.1$	& $0.1 \pm 0.1$	& $0.5 \pm 0.1$\\
Earth$^a$ & 37.8	& 45.0	& 8.05	& 4.45	& 3.55	& 0.36	& 0.25	& 0.01\\
		\hline
{} & \multicolumn{3}{c}{Core composition (wt\%)} & {} &  \multicolumn{3}{c}{Core mass fraction (wt\%)} & {}\\
\cline{2-9} 
{}& Fe & Ni & Si &&&&&\\
$\alpha$-Cen-Earth & $83.3\,_{-7.5}^{+9.0}$	& $5.3\,_{-0.6}^{+0.5}$	& $11.4\,_{-6.3}^{+6.4}$ & {} & \multicolumn{3}{c}{$38.4\,_{-5.1}^{+4.7}$} & {}\\
Earth$^b$ & $82.8\pm2.9$ &	$5.1\pm0.2$ &	$12.1\pm3.0$ & {} & \multicolumn{3}{c}{$32.5\pm0.3$} & {}\\
\hline
\multicolumn{9}{p{17cm}}{$^a$ Refer to the silicate Earth composition (without uncertainties) of \cite{McDonough1995}.}\\
\multicolumn{9}{p{17cm}}{$^b$ Refer to the core composition and core mass fraction of \cite{Wang2018}; silicon composition in the core is rescaled to represent the total estimate of light elements in the core in \cite{Wang2018}.}
	\end{tabular}

\end{table}

For the core composition of a presumed Fe-Ni-Si alloy, the model $\alpha$-Cen-Earth is statistically consistent with Earth. \textbf{However,} the model planet has a relatively higher core mass fraction ($38.4^{+4.7}_{-5.1}$ wt\%) than that of Earth ($32.5\pm0.3$ wt\%; \citealt{Wang2018}). It is noteworthy that, owing to the lack of sulfur abundance measurements in $\alpha$-Cen A and B and the model restrictions on considering other candidate light elements (e.g. C and H), we have only modeled the concentration of Si as the sole light element in the core. Hence, our modeled Si concentration ($11.4^{+6.2}_{-6.5}$ wt\%) should be seen as an upper limit of Si and of light elements, overall.  

\begin{figure*}
	\centering
	\includegraphics[trim=0cm 0cm 0cm 0cm, scale=1.0,angle=0]{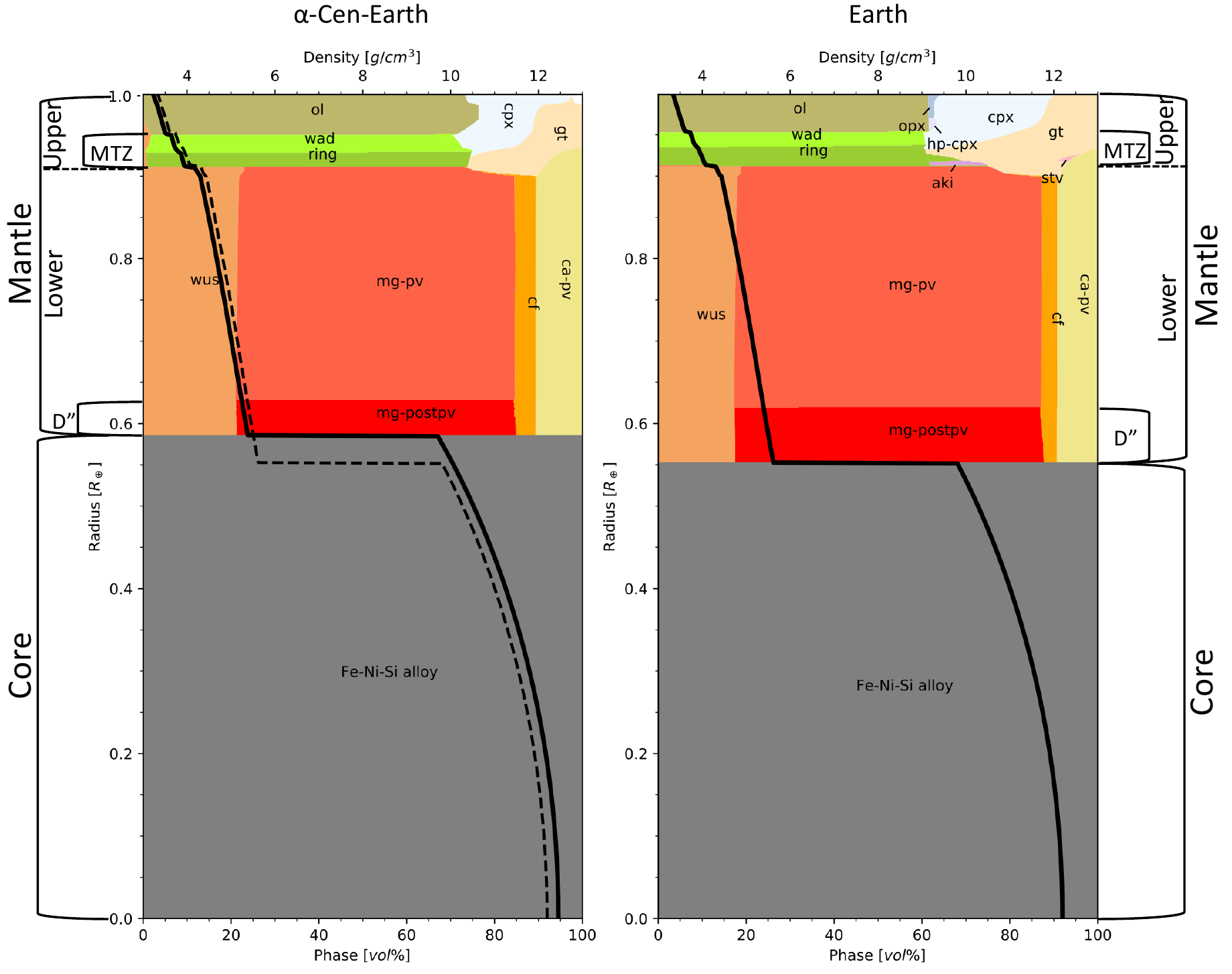} 
	\caption{The comparison of the best-fit mineralogies (in volume fraction, vol\%) and density profiles (solid curves) between the model $\alpha$-Cen-Earth (left panel) and the Earth (right panel). Earth's density profile is replicated as the dashed curve on the left panel for reference. The various layers of the internal structure are noted, including the upper to lower mantle transition zone (MTZ) and the lowermost-mantle layer ($D^{\prime\prime}$). The abbreviated mineral assemblages in the mantle are ol -- olivine, plg -- plagioclase, opx -- orthopyroxene, cpx -- clinopyroxene, hp-cpx -- high-pressure clinopyroxene, wad -- wadsleyite, ring -- ringwoodite, gt -- garnet, aki -- akimotoite, stv -- stishovite, wus -- magnesiowüstite (ferropericlase), mg-pv -- magnesium perovskite (brigdemanite), mg-postpv -- magnesium postperovskite, cf -- calcium-ferrite structured phase, and ca-pv -- calcium perovskite. For more details see Sect. \ref{sec:interior} and Appendix \ref{sec:perplex}.} 
	\label{fig:inteior}
\end{figure*}

With detailed interior modeling results shown in Fig. \ref{fig:inteior}, we find that the model $\alpha$-Cen-Earth has a broadly similar mantle mineralogy \textbf{to} that of the Earth. Specifically, the upper mantle is enriched in olivine (`ol') over pyroxene assemblages (`cpx', `opx', etc.), with the mantle transition zone (MTZ) being dominated by wadsleyite (`wad') and ringwoodite (`ring'). The lower mantle is dominated by the magnesium end-member perovskite (`mg-pv'; (Mg,Fe)SiO$_3$; bridgmanite), except for the lowermost mantle (so-called $D^{\prime\prime}$ zone) that is dominated by the higher-pressure perovskite phase (magnesium post-perovskite, `mg-postpv'). 

In particular, compared to Earth, the olivine/pyroxene ratio (by volume) in the upper mantle of the model $\alpha$-Cen-Earth is equivalent ($2.2\pm1.5$ \textit{vs.} $\sim 1.9$). This highlights the importance of Mg/Si in modulating the dominant mineral assemblages in the upper mantle. 
Likewise, the sum of wadsleyite and ringwoodite in the MTZ is consistent with that of Earth, while the trace phases such as akimotoite (`aki') and stishovite (`stv') vanish in this zone. However, there is a slight increase in the wadsleyite/ringwoodite ratio (by volume) in the model planet relative to Earth ($1.2\pm0.1$ \textit{vs.} $\sim 0.7$). The implications of this will be discussed later. Further, the modeled thickness of the $D^{\prime\prime}$ zone in the lower mantle of the $\alpha$-Cen-Earth is noticeably thinner, with a decreased amount of the high-pressure `mg-postpv' phase ($2.9^{+1.7}_{-1.6}$ \textit{vs.} $\sim 4.9$). This is expected from the slightly enlarged core size and thus elevated core-mantle boundary. A higher Mg\# as revealed above may also suggest that the solid solution in magnesiowüstite (`wus') should be dominated by the periclase (MgO) endmember (instead of FeO).   

Since the lower \textbf{[(O-Mg-2Si)/Fe]} of the planet (relative to Earth) frees a larger fraction of Fe from being oxidized in the mantle ($1.1^{+1.1}_{-0.8}$ wt\% vs. $\sim$ 8 wt\% as for FeO; Table \ref{tab:comp}), thus segregating into the core, the mantle density is slightly but systematically lower and vice-versa for the core density. The relatively elevated core-mantle boundary implies a lower pressure \textbf{at the boundary}, which in turn may lead to less effective Fe$^{2+}$ disproportionation at time of core formation \citep{Frost2004}. 

Importantly, these results should be understood in the context of the yet-large uncertainty of the interior modeling (Fig. \ref{fig:mineral_error}). Overall, the model $\alpha$-Cen-Earth has a broadly Earth-like mineralogy and structure. 
 

\section{Discussion}
\subsection{A reduced primitive mantle and its outgassing}\label{sec:fo2_gases}

The modeled mass fractions of FeO in mantle rocks and Fe in the core (Table \ref{tab:comp}), permit calculation of the model $\alpha$-Cen-Earth's nominal $f\textrm{O}_2$ relative to IW ($\Delta$IW), which presumes equilibrium between the mantle and core:
\begin{equation}
\Delta \textrm{IW} = 2\log(\frac{x^{\textrm{mantle}}_{\textrm{FeO}}}{x^{\textrm{core}}_{\textrm{Fe}}}) + 2\log(\frac{\gamma^{\textrm{mantle}}_{\textrm{FeO}}}{\gamma^{\textrm{core}}_{\textrm{Fe}}}) 
\end{equation}
where $x^{\textrm{mantle}}_{\textrm{FeO}}$ and $x^{\textrm{core}}_{\textrm{Fe}}$ are mole fractions of FeO in mantle rocks and Fe in the core, respectively (converted from their corresponding mass fractions given in Table \ref{tab:comp}) and $\gamma^{\textrm{mantle}}_{\textrm{FeO}}$ and $\gamma^{\textrm{core}}_{\textrm{Fe}}$  are activity coefficients for these components. To facilitate comparison, we set the activity coefficient ratio ($\gamma^{\textrm{mantle}}_{\textrm{FeO}}/\gamma^{\textrm{core}}_{\textrm{Fe}}$) to unity, which is \textbf{expected to be} valid at the high-temperature conditions of the primitive mantle of Earth \citep{Inoue2010, Doyle2019, Sossi2020}. 
As such, $\Delta$IW  depends largely on $x^{\textrm{mantle}}_{\textrm{FeO}}/x^{\textrm{core}}_{\textrm{Fe}}$, which can be \textbf{estimated} for our model planet. 

This leads to an $\Delta$IW value of $-4.0^{+0.6}_{-1.4}$, compared to the terrestrial $\Delta$IW  value of -2.2 upon core formation \citep{Frost2004}. Namely, the primitive mantle of the model $\alpha$-Cen-Earth is $\sim 2$ logarithmic units more reduced than that of Earth. \textbf{We emphasize that the calculated $\Delta$IW value could not be an exact reflection of the planetary fO2, which changes over the depth of the planet and also over time due to planetary internal processes (e.g. disproportionation reactions occurring at high pressures; \citealt{Frost2004}).}

Oxygen fugacity \textbf{at the planetary surface} plays a critical role in determining the composition of the outgassed C-O-H species from a magma ocean prior to and during its crystallization \citep{Hirschmann2012, Ortenzi2020}. Such a magma ocean may have been created in the history of the model $\alpha$-Cen-Earth following extensive heating, presuming it also accreted partially through energetic impacts of precursor planetesimals and planetary embryos, in a manner analogous to the Earth. In order to understand the potential diversity of the early atmospheric compositions produced, its H/C ratio and $f\textrm{O}_2$ are required \citep{Sossi2020}. There is no direct constraint on the H/C ratio of the model $\alpha$-Cen-Earth. Given a stellar nebula of [C/H] = $0.256\pm0.017$ dex (i.e. $\sim$ 2 times higher than the solar C/H; Column 4, Table \ref{tab:starabu}) and a relatively higher C/O (Fig. \ref{fig:star_ratio}), we deduce that H/C in a nascent $\alpha$-Cen-Earth \textbf{should} be considerably lower than that in the early Earth, and hence adopt a value of $\log(\textrm{(H/C)}_{\textrm{BSE}}/2) = 0.53$, where bulk silicate Earth (BSE)'s molar H/C is $\sim$ 6.8 \citep{Hirschmann2009, Marty2012}. \textbf{Oxygen fugacity estimates are adopted from presumed core-mantle equilibration conditions, although oxygen fugacity changes as the planet evolves, and as a function of depth (as observed on Earth). Given these caveats, we consider two end-member scenarios of atmosphere formation on the model $\alpha$-Cen-Earth.} The first presumes the atmospheric H/C is unfractionated from the primitive mantle H/C of the model planet, and in the second, we adopt the model of \cite{Sossi2020} to calculate solubilities of individual gas species using basaltic melts as analogues for the magma ocean, presuming an Earth-like mantle mass (see Sect. \ref{sec:factsage}). \textbf{We note that the importance of H/C over the gross atmospheric chemistry may be worth exploring in further \citep[e.g.][]{Liggins2021}.}

\begin{figure*}
	\centering
	\includegraphics[trim=0cm 0cm 0cm 0cm, scale=1.2,angle=0]{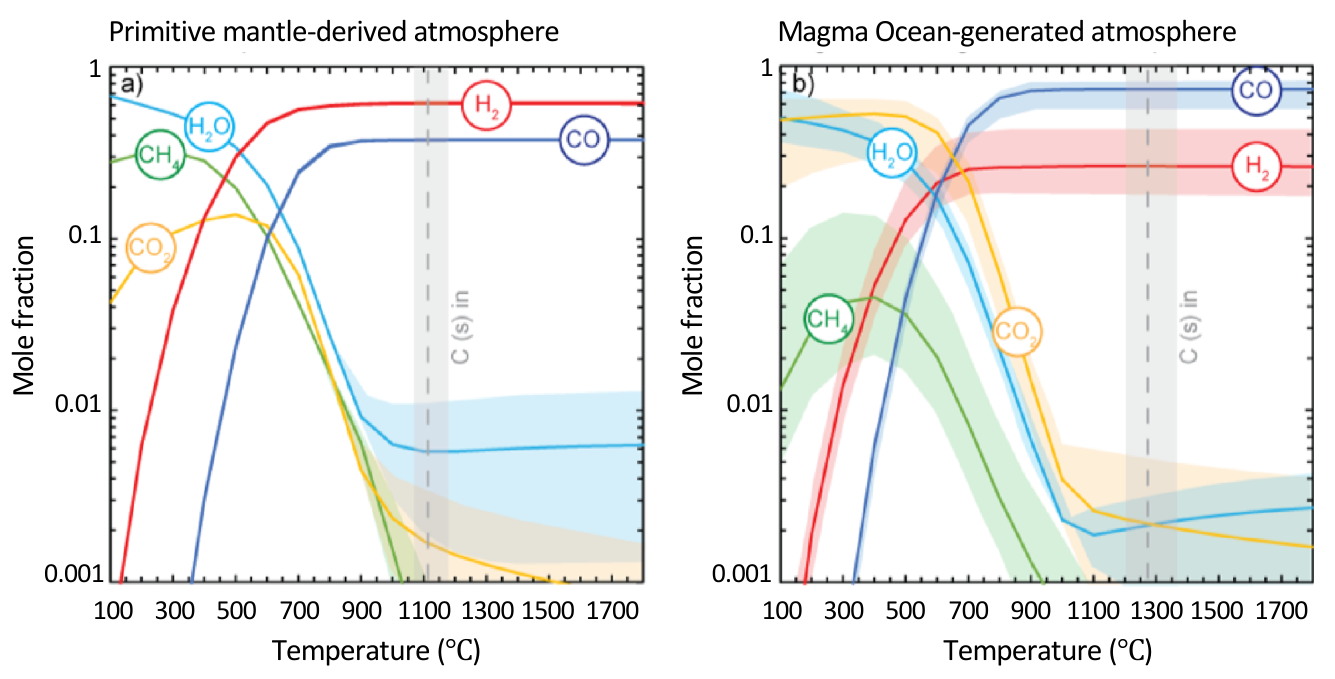} 
	\caption{Calculated model atmospheric compositions, in which species are expressed as mole fractions, for the $\alpha$-Cen-Earth using Gibbs Free Energy minimization in the H-C-O system (see Sect. \ref{sec:factsage}). For the primitive mantle-derived atmosphere (a), the H/C ratio of the atmosphere is presumed to reflect that of the bulk silicate planet, with log(H/C) = 0.53 (molar). For the magma ocean-generated case (b), the H/C ratio of the atmosphere is set by the relative solubilities of the individual gas species in equilibrium with the silicate melt \citep{Sossi2020}. The colored fields represent the range of partial pressures owing to the variation of oxygen fugacity within its uncertainty ($\Delta$IW of -3.4 to -5.4, with the lines showing the case for $\Delta$IW = -4). Note that for the magma ocean-generated case, the log(H/C) of the atmosphere also changes (-0.345, -0.144 and 0.186 at $\Delta$IW = -3.4, -4 -5.4, respectively) because of the lower solubility of H$_2$ relative to H$_2$O, and the increasing H$_2$/H$_2$O ratio as the atmosphere becomes more reducing. The gray vertical dashed line shows the temperature of graphite saturation. For more details see Sect. \ref{sec:fo2_gases} and Appendix \ref{sec:factsage}.} 
	\label{fig:gases}
\end{figure*}

For both cases at magmatic temperatures, the reducing conditions are such that the principal gases are H$_2$ and CO, with the former predominating in high H/C atmospheres (Fig. \ref{fig:gases}a) and vice-versa (Fig. \ref{fig:gases}b). These species are, however, not stable in a cooling atmosphere and react to form CH$_4$ through hydrogenation below about 700 to 800 $^\circ$C leading to an atmosphere dominated by H$_2$O, CH$_4$ and CO$_2$ (Fig. \ref{fig:gases}). Carbon dioxide can be generated through graphite precipitation (H$_2$ + 3CO $\rightarrow$ CO$_2$ + H$_2$O + 2C) and the relative proportions of CO$_2$ to CH$_4$ depend on the $f\textrm{O}_2$ and the H/C in the atmosphere \citep{Sossi2020}. Hence, the primitive mantle-derived case (Fig. \ref{fig:gases}a) produces more CH$_4$ at low temperatures than does the magma ocean-derived scenario (Fig. \ref{fig:gases}b) that is expected to be richer in CO$_2$. This behavior is a consequence of the higher solubility of H relative to C in magmatic liquids (except under extremely reducing conditions; \citealt{Sossi2020, Bower2021}), giving rise to complementary low H/C ratios and thus lower CH$_4$ production rates in magma ocean-generated atmospheres. At $\Delta$IW = -4, about 79 \% of the H budget, but only 1.5 \% of the C budget, is expected to be dissolved in the magma ocean, resulting in a much thinner gaseous envelope than in the nominal ‘primitive mantle’ case in which all H is presumed to be in the atmosphere. Importantly, the fraction of H dissolved scales inversely with $f\textrm{O}_2$, it is $\sim 86$ \% dissolved at $\Delta$IW = -3.5 but only 55 \% at $\Delta$IW = -5.3, owing to the higher H$_2$/H$_2$O ratio in the gas phase and the low solubility of H$_2$ relative to H$_2$O \citep{Hirschmann2012, Moore1998}. Methane is therefore likely to be a ubiquitous gas considering the reduced nature of the model $\alpha$-Cen-Earth mantle. Nonetheless, in both cases, the overall CO$_2$-CH$_4$-H$_2$O-dominated early atmospheres (upon cooling) resemble that of Archean Earth \citep{Gaudi2020, Catling2020}. The subsequent evolution of the Earth, especially by the rising of oxygen 2.5 Gyr ago (known as ``Great Oxidation Event"; \citealt{Lyons2014, Luo2016}) slowly but radically reshaped Earth's atmosphere, which eventually leads to a modern atmosphere dominated by N$_2$ and O$_2$. 
Of course, we have no argument to prescribe a `life-existent' nascent $\alpha$-Cen-Earth, nor can we determine the atmospheric evolution pathway for the planet (beyond what we have modeled for the 'first-generation' early atmospheres). At the extreme and by considering the preferable photolytic destruction to CH$_4$ and H$_2$O \citep{Lasaga1971, Guo2019, Johnstone2020}, we may suspect that a dry, CO$_2$-dominated atmosphere is a likely evolution outcome for such a planet. We will return to this point when we are discussing the stagnant-lid regime for the model planet.   

\subsection{Equivalent water-storage capacity of Earth}
Mantle rocks may be hydrated (if water exists) to different degrees: (i) under relevant upper mantle conditions ($P <$ $\sim 15$ GPa) pyroxene can hold up to 20 times more water in its crystalline structure than olivine \citep{Warren2014}; (ii) the dominant MTZ minerals (wadsleyite \& ringwoodite) can store water in their crystal structures by about one order of magnitude more than any of olivine, pyroxene, and perovskite \citep{Fei2017, Pearson2014, Bercovici2003,  Murakami2002, Bolfan-Casanova2000}; (iii) between wadsleyite and ringwoodite, the former favors water twice as much as the latter based on water partitioning experiments \citep{Inoue2010}. Compared to Earth, the model $\alpha$-Cen-Earth's equivalent amount of wadsleyite + ringwoodite as well as its consistent ratio of olivine/pyroxene dictates that this planet should have a water-storage capacity equivalent to Earth. This capacity is further strengthened by the relatively higher ratio of wadsleyite/ringwoodite ($1.2\pm0.1$ \textit{vs.} $\sim 0.7$). The model $\alpha$-Cen-Earth must therefore be able to store, at least, as much water as in its mantle as Earth. It is worth stressing, however, that ``mantle water-storage capacity" neither implies \textit{presence} of water in a particular abundance \citep[cf.][]{Dorn2015, Unterborn2018, Dencs2019, Shah2021, Acuna2021}, nor does it define the \textit{source} of water \citep[cf.][]{Hallis2015, Peslier2017, Wu2018, OBrien2018, Lichtenberg2019}. A further investigation \textbf{of} these aspects is important for understanding the potential hydrosphere of a rocky planet, but is beyond what can be constrained with the model bulk elemental composition only.  


\subsection{A Venus-like geodynamic regime?}
According to \cite{Wang2020}, an $\alpha$-Cen-Earth with a significantly lower Eu/(Mg+Si) relative to Earth (Fig. \ref{fig:planet_ratio}) has $\sim$ 1/4 and $\sim$ 1/2 less in radiogenic heating from the decay of long-lived radionuclides than Earth at its formation and at present-day, respectively. As calculated in Sect. \ref{sec:fo2_gases}, the primitive mantle of this model planet is also considerably more reduced (by two orders of magnitude) than that of Earth. The former is also modeled to be relatively enriched in graphite/diamond (this should hold still even if some portion of native carbon species has been segregated into the core; \citealt{Dasgupta2013, Li2014, Fichtner2021}), as shown in Table \ref{tab:comp}. Taken together, these factors preferentially impose a stagnant-lid tectonic regime \citep{Hakim2019, Dorn2018, Noack2017, Unterborn2014}. 

In a stagnant-lid regime, volcanism and outgassing of the planet are generally suppressed owing to an effective thermal boundary layer and curtailed convection \citep{Ballmer2021}. For an Earth-sized, stagnant-lid planet, the outgassing of CO$_2$ (up to 50 bar) is permitted, however \citep[][Lena Noack, personal communication]{Dorn2015}. 
Since Fe lowers the solidus temperature of mantle peridodite \citep{Hirschmann2000}, a depleted Fe content in the model $\alpha$-Cen-Earth's mantle (with a significant higher Mg\# of $0.987^{+0.010}_{-0.020}$, versus $0.890 \pm 0.001$ for Earth) further means that partial melting (and hence volcanism) should be less prevalent on the model planet. Over much of the history of the planet, mantle convection and planetary resurfacing should have been moribund. \textbf{This assessment however does not consider the role of melting in heat transport and, particularly under a stagnant lid, heat build-up \citep{Driscoll2014}.} \textbf{Furthermore, a} transient and periodic episode of mobile lids, and thus active volcanism and surging degassing, may take place when accumulated heat beneath a stagnant lid becomes enough to cause catastrophic resurfacing via lithospheric overturn -- i.e. resembling that proposed for the history of Venus: alternating active and stagnant-lid (Turcotte 1993) (cf. localized resurfacing through lava flows; \citealt{Noack2012}). Once the heat has been sufficiently released, however, the mantle \textbf{cools down} and enters into another long-duration episode of stagnant lid tectonics. 

The evolution of planetary heat \textbf{flux} also has important implications for the presence or absence of planetary dynamos. 
Low radiogenic heat production is proposed to favor the generation of geodynamos due to the increased mantle-core temperature difference and thus efficient heat extraction from the core \citep{Nimmo2020}. However, the overlying stagnant-lid regime also ultimately prevents heat extraction from the surface, consequently reducing the mantle-core temperature differences -- i.e. a negative feedback to the generation of geodynamos. The generation and evolution of geodynamos are also related to the planetary rotation \citep{Zuluaga2012} and may also be mechanically stirred by \textbf{stochastic impacts} \citep{Jacobson2017}, all of which are beyond what we can constrain for the $\alpha$-Cen-Earth that is modeled based on our predicted bulk composition only. As such, in spite of a broadly Earth (and potentially Venus)-like interior composition and structure as well as a lower radiogenic heating, it remains an open question if such an $\alpha$-Cen planet would mirror the apparent lack of active dynamos in Venus \citep{Schubert2011, Phillips1987}.  

\section{Conclusions}
Starting with the measured host stellar abundances and a fiducial model of devolatilization, we \textbf{present} an analysis of planetary bulk composition, interiors and (early) atmospheres for a model Earth-sized planet in the habitable zone of $\alpha$-Cen A/B. The detailed analysis offers a new approach of investigation to what we may expect for Earth-sized planets in the habitable zones in the solar neighborhood. Of course, the validity of our analysis and thus conclusion is subject to our principal assumption that our adopted devolatilization model \citep{Wang2019a} may be applied to the $\alpha$ Cen A/B system. Firstly, it is noteworthy that our findings for a HZ rocky planet around the "Sun-like" $\alpha$-Cen A/B should not be extrapolated to its counterpart around Proxima Cen, which is a red dwarf and has a debated origin from the binary companions \citep{Kervella2017, Beech2017, Feng2018}. Second, although $\alpha$ Cen A and B are “Sun-like” stars, their metallicities are $\sim72$\% higher than the solar metallicity (Fig. \ref{fig:ABave}). How this difference would affect the condensation/evaporation process, and thus the devolatilization scale, is the subject of ongoing work \citep{Wang2020a}. Third, we ignore any potential effect of the ``binarity" of the stars on their surrounding planetary bulk chemistry during planet formation, even though we highlight that dynamically, the planetary orbits in the habitable zone around either companion are stable. Finally, we have yet to explore a larger parameter space, e.g. in mass and radius, but have only benchmarked our analysis with an Earth-sized planet, which would otherwise have an impact on the interior modeling \citep[][]{Unterborn2016, Dorn2018, Hinkel2018}. We nonetheless envisage that such an analysis is readily replicable with different parameter settings \textbf{(Wang et al. in preparation)} and is informative as presented to enable such an extension. 

In addition, active research continues over whether a correlation exists between stellar multiplicity on the occurrence rate of planets \citep{Kraus2016, Savel2020}; this extends to the effect of stellar metallicity and the occurrence rate of small/rocky planets \citep{Mulders2016, Petigura2018, Lu2020, Emsenhuber2020}. For the calculation of gas speciation, we presume an H/C ratio for the primitive mantle of the model $\alpha$-Cen-Earth by scaling from the Earth's value with the difference in C/H between $\alpha$-Cen AB and Sun. This approach, however, may be oversimplified. 
Furthermore, we \textbf{do} not consider any catastrophic impact at a scale that may destabilize/remove the mantle or atmosphere of an Earth-sized planet, which may radically alter the geodynamic regimes of the planet including its mantle convection \citep{ONeill2017}, dynamo \citep{Jacobson2017}, and even atmosphere \citep{Lupu2014}. With all of these caveats in mind, we conclude with caution in the following: 
\begin{enumerate}[topsep=4pt, itemsep= 0ex,partopsep=1ex,parsep=1ex]
\item An $\alpha$-Cen-Earth as modeled should have a reduced (primitive) mantle that is similarly dominated by silicates albeit likely enriched in native carbon species (e.g. graphite/diamond); 
\item The planet is also expected to have a slightly larger iron core, with a core mass fraction of $38.4^{+4.7}_{-5.1}$ wt\% (cf. Earth's $32.5\pm0.3$ wt\%);
\item The planet should have an equivalent water-storage capacity of Earth;
\item Such a planet may also have a CO$_2$-CH$_4$-H$_2$O-dominated early atmosphere that resembles that of Archean Earth; observationally, this may be tested with a dry, CO$_2$-dominated atmosphere considering the preferable photolytic destruction to CH$_4$ and H$_2$O and the possibility that \textbf{--}
\item the planet may be in a Venus-like stagnant-lid regime, with sluggish mantle convection and planetary resurfacing, over most of its geological history. 
\end{enumerate}
	


\begin{acknowledgments}
\textbf{We thank the reviewer for useful comments, which helped improve the quality of the manuscript.} We thank Lena Noack, Craig O'Neill, Kaustubh Hakim, Christoph Mordasini, and Yann Alibert for useful discussion. This work has been carried out within the framework of the National Centre of Competence in Research PlanetS supported by the Swiss National Science Foundation (SNSF). H.S.W and S.P.Q acknowledge the financial support of the SNSF. S.J.M. extends special thanks to the University of Vienna, Department of Lithospheric Research (Vienna, Austria) for the Ida Pfeiffer Guest Professorship at which time this manuscript was completed. P.A.S acknowledges SNSF Ambizione Grant 180025. T.M. acknowledges financial support from Belspo for contract PRODEX PLATO mission development. 

\end{acknowledgments}

\pagebreak
\appendix

\section{Software and implementation}\label{sec:software}

\subsection{ExoInt}\label{sec:exoint}
\textit{ExoInt} \citep{Wang2019b} is an open-access package for devolatilizing stellar abundances and computing the mantle composition of major end-member oxides, core composition of an Fe-Ni-Si-S alloy (other candidate light elements such as C and H are not considered yet), as well as core mass fraction. Here we employ an updated version of \textit{ExoInt} (v1.2)\footnote{\url{https://github.com/astro-seanwhy/ExoInt}}, in which the important updates are (i) SiO$_2$ is moved to the end of lithophile oxides and ahead of siderophile oxides (e.g. FeO and NiO), enabling Si to partially sink into the core while the mantle is reduced; (ii) the probability of the valid output from Monte Carlo simulations is computed together with the raw estimate of the mass fraction of each entity in the mantle and core (for details see Table \ref{tab:rawcomp} as well as Appendix A of \citealt{Wang2019b}), and these probabilities are taken as the weighting factors of these raw/unconstrained mass fractions to obtain the final/constrained mantle and core compositions as reported in Table \ref{tab:comp}. 

\begin{table}[!htb]
	\renewcommand\thetable{A1}
	\caption{\label{tab:rawcomp} Unconstrained compositional outputs of the mantle and core and their corresponding probability values ($P$-values). The differences in these $P$-values come from the random draws of different sets of bulk elemental composition with the Monte Carlo simulations ($2\times10^4$ times), which determine two competing scenarios (due to the modeling degeneracy): (i) Si may be partially in the core (i.e. the reduced case where Fe and Ni remain in their metallic phases and 100\% sink into the core); (ii) Si is fully oxidized and thus is not segregated into the core (in such a case Fe and Ni may be partially oxidized in the mantle). As such, the raw outputs of two scenarios are weighted by their $P$-values to obtain the final (constrained) mantle and core compositions (renormalized to 100 wt\%) as reported in Table \ref{tab:comp}. }
	\centering
	\begin{tabular}{c cccc cccc}
		\hline\hline
		{} & \multicolumn{8}{c}{Mantle}\\
		\cline{2-9}
		{}	& MgO	& SiO$_2$	&FeO & Al$_2$O$_3$ &CaO&	Na$_2$O	& NiO	& C\\
		{Composition (wt\%)} & $44.9\,_{-7.0}^{+10.0}$	& $44.0\,_{-9.8}^{+6.4}$&	$8.7\,_{-6.5}^{+10.4}$&	$5.4 \pm 0.7$&	$3.7 \pm 0.5$	& $0.6 \pm 0.1$	& $0.7^{+0.6}_{-0.3}$	& $0.5 \pm 0.1$\\
		{$P$-value} & 1.00 & 1.00 & 0.12 & 1.00 & 1.00 & 1.00& 0.12& 1.00 \\
		\hline
		{} & \multicolumn{6}{c}{Core} &&\\
		\cline{2-7} 
		{} & \multicolumn{2}{c}{Fe} & \multicolumn{2}{c}{Ni} & \multicolumn{2}{c}{Si} &&\\
		Composition (wt\%) & \multicolumn{2}{c}{$83.0\,_{-7.2}^{+9.2}$}	& \multicolumn{2}{c}{$5.2\,_{-0.6}^{+0.5}$}	& \multicolumn{2}{c}{$12.9\,_{-7.3}^{+7.0}$} &&\\
		$P$-value & \multicolumn{2}{c}{1.00}	& \multicolumn{2}{c}{1.00} & \multicolumn{2}{c}{0.88} &&\\
		\hline
	\end{tabular}
\end{table}

\subsection{PerPle$\_$X}\label{sec:perplex}
\begin{figure*}
	\centering
	\renewcommand\thefigure{A1}
	\includegraphics[trim=0cm 0cm 0cm 0cm, scale=0.45,angle=0]{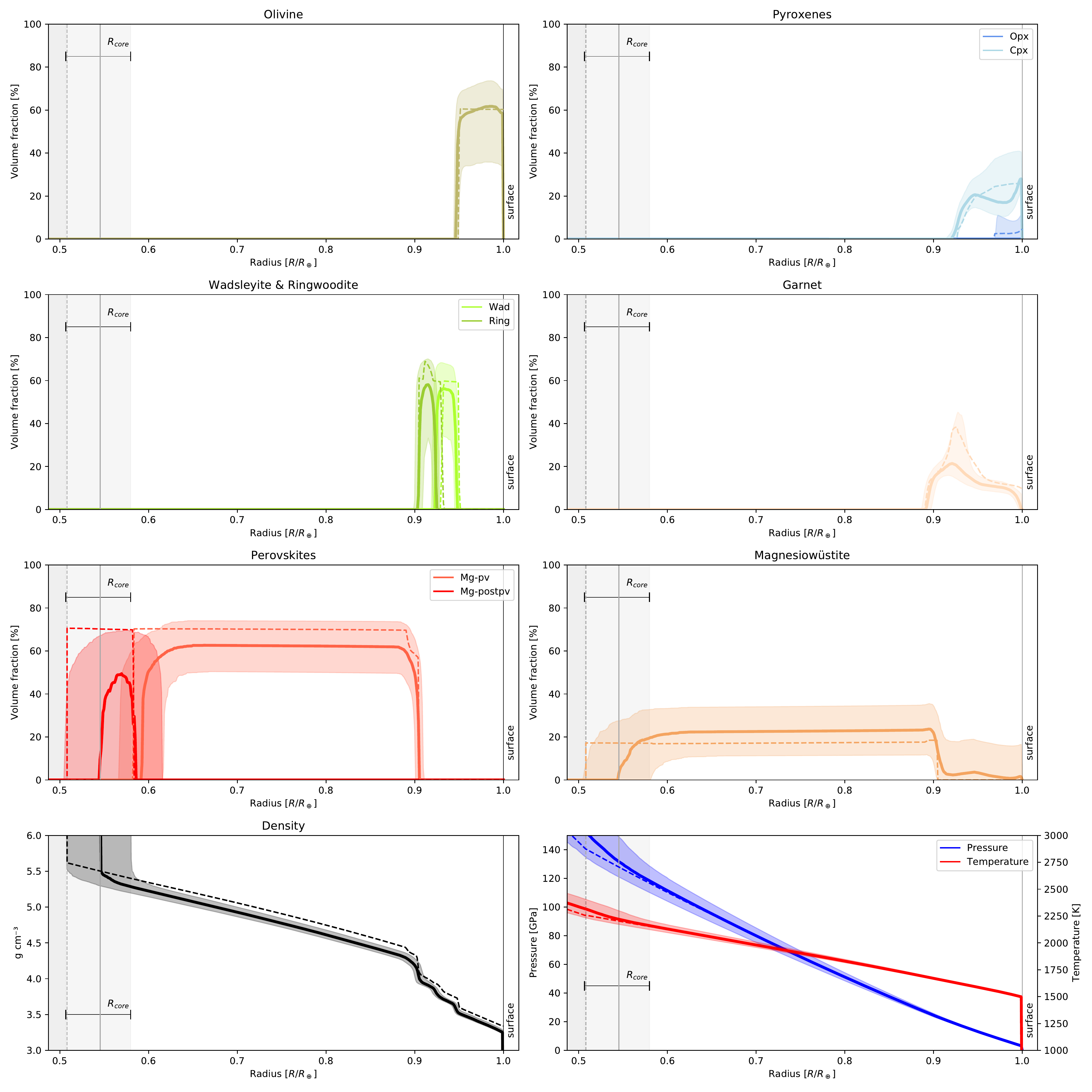} 
	\caption{The uncertainties of the best-fit mineralogy and structure of the model $\alpha$-Cen-Earth. The Earth's best-fit mineralogy as well as density and P-T profiles are shown in dashed curves for comparison. The modeling starts from the base of the lithosphere (i.e. excluding crust/surface). The core regime is partially shown in the gray area towards the left-end of each panel, with the model $\alpha$-Cen-Earth's core radius fraction and its uncertainty labeled as “$R_{\textrm{core}}$” and with Earth's core radius fraction indicated by a vertical dashed line. The data for producing the plot are available upon request.} 
	\label{fig:mineral_error}
\end{figure*}

Perple$\_$X \citep{Connolly2009} is a Gibbs free energy minimization package to compute the mantle mineralogy given the chemical composition and the pressure and temperature gradients. The mineral equations of state (EoS) and thermodynamic parameters of the foremost oxides (SiO$_2$, MgO, FeO, Al$_2$O$_3$, CaO and Na$_2$O) are adopted from \cite{Stixrude2011}. Other less important oxides (e.g. NiO, SO$_3$, and CO$_2$) and all reduced phases (e.g. graphite/diamond and other mantle metals), which are also computable with \textit{ExoInt}, are neglected for the mineralogy modeling. The liquid EoS \citep{Anderson1994} is adopted for an Fe-Ni alloy core (with/without light elements). We adopt an adiabatic thermal gradient as commonly practiced in the literature for rocky exoplanets \citep{Dorn2015, Unterborn2018, Lorenzo2018}. Integration of the gradient starts at the base of the lithosphere (i.e. the top of the uppermost mantle), with an initial pressure and temperature arbitrarily set at 3 Gpa and 1450 K  -- similar to the conditions at Earth's crust-mantle boundary \citep{Anderson1982, Dziewonski1981}. Then the best-fit values of our estimates of the mantle and core compositions and core mass fraction with \textit{ExoInt} for the model $\alpha$-Cen-Earth (Table \ref{tab:comp}), along with the assumption of 1$M_{\oplus}$ and $R_{\oplus}$, are input to \textit{Perple$\_$X} to obtain the self-consistent mineralogy and density profile as shown in Fig. \ref{fig:inteior}. Monte Carlo simulations with the random values drawn from our mantle and core compositional estimates (Table \ref{tab:comp}) are carried out to obtain the uncertainties of individual minerals as shown in Fig. \ref{fig:mineral_error}. For generating the best-fit mineralogy for Earth, the input mantle and core compositions are from Table \ref{tab:comp} (excluding NiO and C as well as the uncertainties associated with the core compositions). 

\subsection{FactSage 8.0}\label{sec:factsage}
\textit{FactSage 8.0} \citep{Bale2016} is employed to calculate the gas speciation as a function of temperature. The composition used is set in the H-C-O system for a real gas and pure solid and liquid phases, and a nominal total pressure of 1 bar. The mole fraction of O adjusted to yield the desired relative oxygen fugacity at 2073 K. The calculations were then stepped in 100 K temperature intervals down to 373 K and the equilibrium speciation and any condensing solid or liquid phases resolved. In the magma ocean-generated cases, the primitive mantle's H/C was subjected to partitioning between a magma ocean and its atmosphere at 2073 K, in order to calculate a new H/C ratio of the atmosphere. To do so, the H abundance was fixed at the value of Earth's mantle (0.012 wt\%) \citep{Palme2014b}, and the C content increased to 0.043 wt\% to yield the desired molar log(H/C) ratio (0.53; see main text). Partial pressures were then resolved according to mass balance given the mass of the mantle (assumed to be that of the Earth's, $4.01\times10^{24}$ kg) and the acceleration due to gravity. The mass balance must also satisfy the condition of redox equilibrium, meaning that the H$_2$/H$_2$O and CO/CO$_2$ ratios are uniquely determined by the fixed temperature, pressure and oxygen fugacity. The partial pressure of each species was then iteratively determined by considering the solubility of H$_2$O, H$_2$, CO, and CO$_2$ in basaltic melts as proxies for the magma ocean (for details see \citealt{Sossi2020} and references therein). 

\bibliographystyle{aasjournal}
\bibliography{MyPapersBib}

\begin{thebibliography}{}
\expandafter\ifx\csname natexlab\endcsname\relax\def\natexlab#1{#1}\fi
\providecommand{\url}[1]{\href{#1}{#1}}
\providecommand{\dodoi}[1]{doi:~\href{http://doi.org/#1}{\nolinkurl{#1}}}
\providecommand{\doeprint}[1]{\href{http://ascl.net/#1}{\nolinkurl{http://ascl.net/#1}}}
\providecommand{\doarXiv}[1]{\href{https://arxiv.org/abs/#1}{\nolinkurl{https://arxiv.org/abs/#1}}}

\bibitem[{Acu{\~{n}}a(2019)}]{Acuna2019}
Acu{\~{n}}a, L. 2019, Master's thesis, Lund University.
\newblock \url{https://www.lunduniversity.lu.se/lup/publication/8976827}

\bibitem[{Acuna {et~al.}(2021)Acuna, Deleuil, Mousis, Marcq, Levesque, \&
  Aguichine}]{Acuna2021}
Acuna, L., Deleuil, M., Mousis, O., {et~al.} 2021, Astron. Astrophys., 647, 1,
  \dodoi{10.1051/0004-6361/202039885}

\bibitem[{Adibekyan {et~al.}(2021)Adibekyan, Dorn, Sousa, Santos, Bitsch,
  Israelian, Mordasini, Barros, {Delgado Mena}, Demangeon, Faria, Figueira,
  Hakobyan, Oshagh, Soares, Kunitomo, Takeda, Jofr{\'{e}}, Petrucci, \&
  Martioli}]{Adibekyan2021}
Adibekyan, V., Dorn, C., Sousa, S.~G., {et~al.} 2021, Science (80-. )., 374,
  330, \dodoi{10.1126/science.abg8794}

\bibitem[{Akeson {et~al.}(2021)Akeson, Beichman, Kervella, Fomalont, \&
  Benedict}]{Akeson2021}
Akeson, R., Beichman, C., Kervella, P., Fomalont, E., \& Benedict, G.~F. 2021.
\newblock \doarXiv{2104.10086}

\bibitem[{Anderson(1982)}]{Anderson1982}
Anderson, O. 1982, Phil. Trans. R. So. Lond. A, 306, 21

\bibitem[{Anderson \& Ahrens(1994)}]{Anderson1994}
Anderson, W., \& Ahrens, T. 1994, J. Geophys. Res., 99, 4273

\bibitem[{Andrade-Ines \& Michtchenko(2014)}]{Andrade2014}
Andrade-Ines, E., \& Michtchenko, T.~A. 2014, Mon. Not. R. Astron. Soc., 444,
  2167, \dodoi{10.1093/mnras/stu1591}

\bibitem[{Anglada-Escud{\'{e}} {et~al.}(2016)Anglada-Escud{\'{e}}, Amado,
  Barnes, Berdi{\~{n}}as, Butler, Coleman, de~la Cueva, Dreizler, Endl,
  Giesers, Jeffers, Jenkins, Jones, Kiraga, K{\"{u}}rster,
  L{\'{o}}pez-Gonz{\'{a}}lez, Marvin, Morales, Morin, Nelson, Ortiz, Ofir,
  Paardekooper, Reiners, Rodr{\'{i}}guez, Rodrίguez-L{\'{o}}pez, Sarmiento,
  Strachan, Tsapras, Tuomi, \& Zechmeister}]{Anglada2016}
Anglada-Escud{\'{e}}, G., Amado, P.~J., Barnes, J., {et~al.} 2016, Nature, 536,
  437, \dodoi{10.1038/nature19106}

\bibitem[{Armstrong {et~al.}(2019)Armstrong, Frost, McCammon, Rubie, \&
  Ballaran}]{Armstrong2019}
Armstrong, K., Frost, D.~J., McCammon, C.~A., Rubie, D.~C., \& Ballaran, T.~B.
  2019, Science (80-. )., 365, 903, \dodoi{10.1126/science.aax8376}

\bibitem[{Asplund {et~al.}(2021)Asplund, Amarsi, \& Grevesse}]{Asplund2021}
Asplund, M., Amarsi, A., \& Grevesse, N. 2021, Astron. Astrophys., 1,
  \dodoi{10.1051/0004-6361/202140445}

\bibitem[{Bale {et~al.}(2016)Bale, B{\'{e}}lisle, Chartrand, Decterov,
  Eriksson, Gheribi, Hack, Jung, Kang, Melan{\c{c}}on, Pelton, Petersen,
  Robelin, Sangster, Spencer, \& {Van Ende}}]{Bale2016}
Bale, C.~W., B{\'{e}}lisle, E., Chartrand, P., {et~al.} 2016, Calphad Comput.
  Coupling Phase Diagrams Thermochem., 55, 1,
  \dodoi{10.1016/j.calphad.2016.07.004}

\bibitem[{Ballhaus {et~al.}(1990)Ballhaus, Berry, \& Green}]{Ballhaus1990}
Ballhaus, C., Berry, R.~F., \& Green, D.~H. 1990, Nature, 348, 437,
  \dodoi{10.1038/348437a0}

\bibitem[{Ballmer \& Noack(2021)}]{Ballmer2021}
Ballmer, M.~D., \& Noack, L. 2021.
\newblock \doarXiv{2108.08385}

\bibitem[{Beech(2015)}]{Beech2015}
Beech, M. 2015, {Alpha Centauri: unveiling the secrets of our nearest stellar
  neighbor} (Springer).
\newblock \url{https://www.springer.com/de/book/9783319093710}

\bibitem[{Beech {et~al.}(2017)Beech, McCowan, \& Peltier}]{Beech2017}
Beech, M., McCowan, C., \& Peltier, L. 2017, Am. J. Astron. Astrophys., 5, 1,
  \dodoi{10.11648/j.ajaa.20170501.11}

\bibitem[{Beichman {et~al.}(2020)Beichman, Ygouf, Sayson, Mawet, Yung, Choquet,
  Kervella, Boccaletti, Belikov, Lissauer, Quarles, Lagage, Dicken, Hu,
  Mennesson, Ressler, Serabyn, Krist, Bendek, Leisenring, \&
  Pueyo}]{Beichman2020}
Beichman, C., Ygouf, M., Sayson, J.~L., {et~al.} 2020, Publ. Astron. Soc.
  Pacific, 132, 15002, \dodoi{10.1088/1538-3873/ab5066}

\bibitem[{Bercovici \& Karato(2003)}]{Bercovici2003}
Bercovici, D., \& Karato, S.~I. 2003, Nature, 425, 39,
  \dodoi{10.1038/nature01918}

\bibitem[{Bolfan-Casanova {et~al.}(2000)Bolfan-Casanova, Keppler, \&
  Rubie}]{Bolfan-Casanova2000}
Bolfan-Casanova, N., Keppler, H., \& Rubie, D.~C. 2000, Earth Planet. Sci.
  Lett., 182, 209, \dodoi{10.1016/S0012-821X(00)00244-2}

\bibitem[{Bond {et~al.}(2010)Bond, O'Brien, \& Lauretta}]{Bond2010}
Bond, J.~C., O'Brien, D.~P., \& Lauretta, D.~S. 2010, Astrophys. J., 715, 1050,
  \dodoi{10.1088/0004-637X/715/2/1050}

\bibitem[{Bowens {et~al.}(2021)Bowens, Meyer, Delacroix, Absil, van Boekel,
  Quanz, Shinde, Kenworthy, Carlomagno, {Orban de Xivry}, Cantalloube, \&
  Pathak}]{Bowens2021}
Bowens, R., Meyer, M.~R., Delacroix, C., {et~al.} 2021, Astron. Astrophys.,
  653, A8, \dodoi{10.1051/0004-6361/202141109}

\bibitem[{Bower {et~al.}(2021)Bower, Hakim, Sossi, \& Sanan}]{Bower2021}
Bower, D.~J., Hakim, K., Sossi, P.~A., \& Sanan, P. 2021.
\newblock \doarXiv{2110.08029}

\bibitem[{Brewer \& Fischer(2016)}]{Brewer2016}
Brewer, J.~M., \& Fischer, D.~A. 2016, Astrophys. J., 831, 1,
  \dodoi{10.3847/0004-637X/831/1/20}

\bibitem[{Bryson {et~al.}(2021)Bryson, Kunimoto, Kopparapu, Coughlin, Borucki,
  Koch, Aguirre, Allen, Barentsen, Batalha, Berger, Boss, Buchhave, Burke,
  Caldwell, Campbell, Catanzarite, Chandrasekaran, Chaplin, Christiansen,
  Christensen-Dalsgaard, Ciardi, Clarke, Cochran, Dotson, Doyle, Duarte,
  Dunham, Dupree, Endl, Fanson, Ford, Fujieh, {Gautier III}, Geary, Gilliland,
  Girouard, Gould, Haas, Henze, Holman, Howard, Howell, Huber, Hunter, Jenkins,
  Kjeldsen, Kolodziejczak, Larson, Latham, Li, Mathur, Meibom, Middour, Morris,
  Morton, Mullally, Mullally, Pletcher, Prsa, Quinn, Quintana, Ragozzine,
  Ramirez, Sanderfer, Sasselov, Seader, Shabram, Shporer, Smith, Steffen,
  Still, Torres, Troeltzsch, Twicken, Uddin, {Van Cleve}, Voss, Weiss, Welsh,
  Wohler, \& Zamudio}]{Bryson2021}
Bryson, S., Kunimoto, M., Kopparapu, R.~K., {et~al.} 2021, Astron. J., 161, 36,
  \dodoi{10.3847/1538-3881/abc418}

\bibitem[{Carlomagno(2020)}]{Carlomagno2020}
Carlomagno, B. 2020, J. Astron. Telesc. Instruments, Syst., 6, 1,
  \dodoi{10.1117/1.JATIS.6.3.035005}

\bibitem[{Carlson {et~al.}(2014)Carlson, Garnero, Harrison, Li, Manga,
  McDonough, Mukhopadhyay, Romanowicz, Rubie, Williams, \& Zhong}]{Carlson2014}
Carlson, R.~W., Garnero, E., Harrison, T.~M., {et~al.} 2014, Annu. Rev. Earth
  Planet. Sci., 42, 151, \dodoi{10.1146/annurev-earth-060313-055016}

\bibitem[{Cassan {et~al.}(2012)Cassan, Kubas, Beaulieu, Dominik, Horne,
  Greenhill, Wambsganss, Menzies, Williams, Jorgensen, Udalski, Bennett,
  Albrow, Batista, Brillant, Caldwell, Cole, Coutures, Cook, Dieters, Prester,
  Donatowicz, Fouqu{\'{e}}, Hill, Kains, Kane, Marquette, Martin, Pollard,
  Sahu, Vinter, Warren, Watson, Zub, Sumi, Szyma{\'{n}}ski, Kubiak, Poleski,
  Soszynski, Ulaczyk, Pietrzy{\'{n}}ski, \& Wyrzykowski}]{Cassan2012}
Cassan, A., Kubas, D., Beaulieu, J.~P., {et~al.} 2012, Nature, 481, 167,
  \dodoi{10.1038/nature10684}

\bibitem[{Catling \& Zahnle(2020)}]{Catling2020}
Catling, D.~C., \& Zahnle, K.~J. 2020, Sci. Adv., 6,
  \dodoi{10.1126/sciadv.aax1420}

\bibitem[{Clark {et~al.}(2021)Clark, Clert{\'{e}}, Hinkel, Unterborn,
  Wittenmyer, Horner, Wright, Carter, Morton, Spina, Asplund, Buder,
  Bland-Hawthorn, Casey, {De Silva}, D'Orazi, Duong, Hayden, Freeman, Kos,
  Lewis, Lin, Lind, Martell, Sharma, Simpson, Zucker, Zwitter, Tinney, Ting,
  Nordlander, \& Amarsi}]{Clark2021}
Clark, J.~T., Clert{\'{e}}, M., Hinkel, N.~R., {et~al.} 2021, Mon. Not. R.
  Astron. Soc., 504, 4968, \dodoi{10.1093/mnras/stab1052}

\bibitem[{Connolly(2009)}]{Connolly2009}
Connolly, J. A.~D. 2009, Geochemistry, Geophys. Geosystems, 10,
  \dodoi{10.1029/2009GC002540}

\bibitem[{Cowley {et~al.}(2021)Cowley, Bord, \& Y{\"{u}}ce}]{Cowley2021}
Cowley, C.~R., Bord, D.~J., \& Y{\"{u}}ce, K. 2021, Astron. J., 161, 142,
  \dodoi{10.3847/1538-3881/abdf5d}

\bibitem[{Damasso {et~al.}(2020)Damasso, Sordo, Anglada-Escud{\'{e}}, Giacobbe,
  Sozzetti, Morbidelli, Pojmanski, Barbato, {Paul Butler}, Jones, Hambsch,
  Jenkins, L{\'{o}}pez-Gonz{\'{a}}lez, Morales, {Pe{\~{n}}a Rojas},
  Rodr{\'{i}}guez-L{\'{o}}pez, Rodr{\'{i}}guez, Amado, Anglada, Feng, \&
  G{\'{o}}mez}]{Damasso2020}
Damasso, M., Sordo, F.~D., Anglada-Escud{\'{e}}, G., {et~al.} 2020, Sci. Adv.,
  6, 1, \dodoi{10.1126/sciadv.aax7467}

\bibitem[{Dasgupta {et~al.}(2013)Dasgupta, Chi, Shimizu, Buono, \&
  Walker}]{Dasgupta2013}
Dasgupta, R., Chi, H., Shimizu, N., Buono, A.~S., \& Walker, D. 2013, Geochim.
  Cosmochim. Acta, 102, 191, \dodoi{10.1016/j.gca.2012.10.011}

\bibitem[{{Delgado Mena} {et~al.}(2010){Delgado Mena}, Israelian,
  {Gonz{\'{a}}lez Hern{\'{a}}ndez}, Bond, Santos, Udry, \& Mayor}]{Delgado2010}
{Delgado Mena}, E., Israelian, G., {Gonz{\'{a}}lez Hern{\'{a}}ndez}, J.~I.,
  {et~al.} 2010, Astrophys. J., 725, 2349, \dodoi{10.1088/0004-637X/725/2/2349}

\bibitem[{Dencs \& Reg{\'{a}}ly(2019)}]{Dencs2019}
Dencs, Z., \& Reg{\'{a}}ly, Z. 2019, Mon. Not. R. Astron. Soc., 487, 2191,
  \dodoi{10.1093/mnras/stz1412}

\bibitem[{Dorn {et~al.}(2018)Dorn, Bower, \& Rozel}]{Dorn2018}
Dorn, C., Bower, D.~J., \& Rozel, A. 2018, in Handb. Exopl. (Cham: Springer
  International Publishing), 1--25, \dodoi{10.1007/978-3-319-30648-3_66-1}

\bibitem[{Dorn {et~al.}(2019)Dorn, Harrison, Bonsor, \& Hands}]{Dorn2019}
Dorn, C., Harrison, J.~H., Bonsor, A., \& Hands, T.~O. 2019, Mon. Not. R.
  Astron. Soc., 484, 712, \dodoi{10.1093/mnras/sty3435}

\bibitem[{Dorn {et~al.}(2015)Dorn, Khan, Heng, Connolly, Alibert, Benz, \&
  Tackley}]{Dorn2015}
Dorn, C., Khan, A., Heng, K., {et~al.} 2015, Astron. Astrophys., 577, A83,
  \dodoi{10.1051/0004-6361/201424915}

\bibitem[{Doyle {et~al.}(2019)Doyle, Young, Klein, Zuckerman, \&
  Schlichting}]{Doyle2019}
Doyle, A.~E., Young, E.~D., Klein, B., Zuckerman, B., \& Schlichting, H.~E.
  2019, Science, 366, 356, \dodoi{10.1126/science.aax3901}

\bibitem[{Driscoll \& Bercovici(2014)}]{Driscoll2014}
Driscoll, P., \& Bercovici, D. 2014, Phys. Earth Planet. Inter., 236, 36,
  \dodoi{10.1016/j.pepi.2014.08.004}

\bibitem[{Dumusque {et~al.}(2012)Dumusque, Pepe, Lovis, S{\'{e}}gransan,
  Sahlmann, Benz, Bouchy, Mayor, Queloz, Santos, \& Udry}]{Dumusque2012}
Dumusque, X., Pepe, F., Lovis, C., {et~al.} 2012, Nature, 1, 4,
  \dodoi{10.1038/nature11572}

\bibitem[{Dziewonski \& Anderson(1981)}]{Dziewonski1981}
Dziewonski, A.~M., \& Anderson, D.~L. 1981, Phys. Earth Planet. Inter., 25,
  297, \dodoi{10.1016/0031-9201(81)90046-7}

\bibitem[{Emsenhuber {et~al.}(2020)Emsenhuber, Mordasini, Burn, Alibert, Benz,
  \& Asphaug}]{Emsenhuber2020}
Emsenhuber, A., Mordasini, C., Burn, R., {et~al.} 2020.
\newblock \doarXiv{2007.05562}

\bibitem[{Fegley {et~al.}(2020)Fegley, Lodders, \& Jacobson}]{Fegley2020}
Fegley, B., Lodders, K., \& Jacobson, N.~S. 2020, Chemie der Erde, 80, 125594,
  \dodoi{10.1016/j.chemer.2019.125594}

\bibitem[{Fei {et~al.}(2017)Fei, Yamazaki, Sakurai, Miyajima, Ohfuji, Katsura,
  \& Yamamoto}]{Fei2017}
Fei, H., Yamazaki, D., Sakurai, M., {et~al.} 2017, Sci. Adv., 3, 1,
  \dodoi{10.1126/sciadv.1603024}

\bibitem[{Feng \& Jones(2018)}]{Feng2018}
Feng, F., \& Jones, H.~R. 2018, Mon. Not. R. Astron. Soc., 473, 3185,
  \dodoi{10.1093/mnras/stx2576}

\bibitem[{Fichtner {et~al.}(2021)Fichtner, Schmidt, Liebske, Bouvier, \&
  Baumgartner}]{Fichtner2021}
Fichtner, C.~E., Schmidt, M.~W., Liebske, C., Bouvier, A.~S., \& Baumgartner,
  L.~P. 2021, Earth Planet. Sci. Lett., 554, 116659,
  \dodoi{10.1016/j.epsl.2020.116659}

\bibitem[{Fortney(2012)}]{Fortney2012}
Fortney, J.~J. 2012, Astrophys. J., 747, L27,
  \dodoi{10.1088/2041-8205/747/2/L27}

\bibitem[{Frank {et~al.}(2014)Frank, Meyer, \& Mojzsis}]{Frank2014}
Frank, E.~a., Meyer, B.~S., \& Mojzsis, S.~J. 2014, Icarus, 243, 274,
  \dodoi{10.1016/j.icarus.2014.08.031}

\bibitem[{Frost {et~al.}(2004)Frost, Liebske, Langenhorst, McCammon,
  Tr{\o}nnes, \& Rubie}]{Frost2004}
Frost, D.~J., Liebske, C., Langenhorst, F., {et~al.} 2004, Nature, 428, 409,
  \dodoi{10.1038/nature02413}

\bibitem[{Gastis {et~al.}(2020)Gastis, Perdikakis, Dissanayake, Tsintari,
  Sultana, Brune, Massey, Meisel, Voinov, Brandenburg, Danley, Giri,
  Jones-Alberty, Paneru, Soltesz, \& Subedi}]{Gastis2020}
Gastis, P., Perdikakis, G., Dissanayake, J., {et~al.} 2020, arXiv.
\newblock \doarXiv{2001.11600}

\bibitem[{Gaudi {et~al.}(2020)Gaudi, Seager, Mennesson, Kiessling, Warfield,
  Cahoy, Clarke, Domagal-Goldman, Feinberg, Guyon, Kasdin, Mawet, Plavchan,
  Robinson, Rogers, Scowen, Somerville, Stapelfeldt, Stark, Stern, Turnbull,
  Amini, Kuan, Martin, Morgan, Redding, Stahl, Webb, Alvarez-Salazar, Arnold,
  Arya, Balasubramanian, Baysinger, Bell, Below, Benson, Blais, Booth,
  Bourgeois, Bradford, Brewer, Brooks, Cady, Caldwell, Calvet, Carr, Chan,
  Cormarkovic, Coste, Cox, Danner, Davis, Dewell, Dorsett, Dunn, East,
  Effinger, Eng, Freebury, Garcia, Gaskin, Greene, Hennessy, Hilgemann, Hood,
  Holota, Howe, Huang, Hull, Hunt, Hurd, Johnson, Kissil, Knight, Kolenz,
  Kraus, Krist, Li, Lisman, Mandic, Mann, Marchen, Marrese-Reading, McCready,
  McGown, Missun, Miyaguchi, Moore, Nemati, Nikzad, Nissen, Novicki, Perrine,
  Pineda, Polanco, Putnam, Qureshi, Richards, Riggs, Rodgers, Rud, Saini,
  Scalisi, Scharf, Schulz, Serabyn, Sigrist, Sikkia, Singleton, Shaklan, Smith,
  Southerd, Stahl, Steeves, Sturges, Sullivan, Tang, Taras, Tesch, Therrell,
  Tseng, Valente, {Van Buren}, Villalvazo, Warwick, Webb, Westerhoff, Wofford,
  Wu, Woo, Wood, Ziemer, Arney, Anderson, Ma{\'{i}}z-Apell{\'{a}}niz, Bartlett,
  Belikov, Bendek, Cenko, Douglas, Dulz, Evans, Faramaz, Feng, Ferguson,
  Follette, Ford, Garc{\'{i}}a, Geha, Gelino, G{\"{o}}tberg, Hildebrandt, Hu,
  Jahnke, Kennedy, Kreidberg, Isella, Lopez, Marchis, Macri, Marley, Matzko,
  Mazoyer, McCandliss, Meshkat, Mordasini, Morris, Nielsen, Newman, Petigura,
  Postman, Reines, Roberge, Roederer, Ruane, Schwieterman, Sirbu, Spalding,
  Teplitz, Tumlinson, Turner, Werk, Wofford, Wyatt, Young, \&
  Zellem}]{Gaudi2020}
Gaudi, B.~S., Seager, S., Mennesson, B., {et~al.} 2020, {The Habitable
  Exoplanet Observatory (HabEx) Mission Concept Study Final Report}, Tech. rep.
\newblock \doarXiv{2001.06683}

\bibitem[{Grossman \& Larimer(1974)}]{Grossman1974}
Grossman, L., \& Larimer, J.~W. 1974, Rev. Geophys., 12, 71,
  \dodoi{10.1029/RG012i001p00071}

\bibitem[{Guiglion {et~al.}(2018)Guiglion, {De Laverny}, Recio-Blanco, \&
  Prantzos}]{Guiglion2018}
Guiglion, G., {De Laverny}, P., Recio-Blanco, A., \& Prantzos, N. 2018, Astron.
  Astrophys., 619, 1, \dodoi{10.1051/0004-6361/201833782}

\bibitem[{Guo(2019)}]{Guo2019}
Guo, J.~H. 2019, Astrophys. J., 872, 99, \dodoi{10.3847/1538-4357/aaffd4}

\bibitem[{Hakim {et~al.}(2019)Hakim, Spaargaren, Grewal, Rohrbach, Berndt,
  Dominik, \& {Van Westrenen}}]{Hakim2019}
Hakim, K., Spaargaren, R., Grewal, D.~S., {et~al.} 2019, Astrobiology, 19, 867,
  \dodoi{10.1089/ast.2018.1930}

\bibitem[{Hallis {et~al.}(2015)Hallis, Huss, Nagashima, Taylor,
  Halld{\'{o}}rsson, Hilton, Mottl, \& Meech}]{Hallis2015}
Hallis, L.~J., Huss, G.~R., Nagashima, K., {et~al.} 2015, Science (80-. ).,
  350, 795, \dodoi{10.1126/science.aac4834}

\bibitem[{Harrison {et~al.}(2018)Harrison, Bonsor, \&
  Madhusudhan}]{Harrison2018}
Harrison, J.~H., Bonsor, A., \& Madhusudhan, N. 2018, Mon. Not. R. Astron.
  Soc., 479, 3814, \dodoi{10.1093/mnras/sty1700}

\bibitem[{Harrison {et~al.}(2021)Harrison, Shorttle, \& Bonsor}]{Harrison2021}
Harrison, J.~H., Shorttle, O., \& Bonsor, A. 2021, Earth Planet. Sci. Lett.,
  554, 116694, \dodoi{10.1016/j.epsl.2020.116694}

\bibitem[{Heiter {et~al.}(2015)Heiter, Jofr{\'{e}}, Gustafsson, Korn, Soubiran,
  \& Th{\'{e}}venin}]{Heiter2015}
Heiter, U., Jofr{\'{e}}, P., Gustafsson, B., {et~al.} 2015, Astron. Astrophys.,
  582, 1, \dodoi{10.1051/0004-6361/201526319}

\bibitem[{Hin {et~al.}(2017)Hin, Coath, Carter, Nimmo, Lai, {Pogge von
  Strandmann}, Willbold, Leinhardt, Walter, \& Elliott}]{Hin2017}
Hin, R.~C., Coath, C.~D., Carter, P.~J., {et~al.} 2017, Nature, 549, 511,
  \dodoi{10.1038/nature23899}

\bibitem[{Hinkel \& Unterborn(2018)}]{Hinkel2018}
Hinkel, N.~R., \& Unterborn, C.~T. 2018, Astrophys. J., 853, 83,
  \dodoi{10.3847/1538-4357/aaa5b4}

\bibitem[{Hirschmann(2000)}]{Hirschmann2000}
Hirschmann, M.~M. 2000, Geochemistry, Geophys. Geosystems, 1,
  \dodoi{10.1029/2000GC000070}

\bibitem[{Hirschmann(2012)}]{Hirschmann2012}
---. 2012, Earth Planet. Sci. Lett., 341-344, 48,
  \dodoi{10.1016/j.epsl.2012.06.015}

\bibitem[{Hirschmann \& Dasgupta(2009)}]{Hirschmann2009}
Hirschmann, M.~M., \& Dasgupta, R. 2009, Chem. Geol., 262, 4,
  \dodoi{10.1016/j.chemgeo.2009.02.008}

\bibitem[{Inoue {et~al.}(2010)Inoue, Wada, Sasaki, \& Yurimoto}]{Inoue2010}
Inoue, T., Wada, T., Sasaki, R., \& Yurimoto, H. 2010, Phys. Earth Planet.
  Inter., 183, 245, \dodoi{10.1016/j.pepi.2010.08.003}

\bibitem[{Jacobson {et~al.}(2017)Jacobson, Rubie, Hernlund, Morbidelli, \&
  Nakajima}]{Jacobson2017}
Jacobson, S.~A., Rubie, D.~C., Hernlund, J., Morbidelli, A., \& Nakajima, M.
  2017, Earth Planet. Sci. Lett., 474, 375, \dodoi{10.1016/j.epsl.2017.06.023}

\bibitem[{Jofr{\'{e}} {et~al.}(2017)Jofr{\'{e}}, Das, Bertranpetit, \&
  Foley}]{Jofre2017}
Jofr{\'{e}}, P., Das, P., Bertranpetit, J., \& Foley, R. 2017, Mon. Not. R.
  Astron. Soc., 467, 1140, \dodoi{10.1093/mnras/stx075}

\bibitem[{Johnstone(2020)}]{Johnstone2020}
Johnstone, C.~P. 2020, Astrophys. J., 890, 79, \dodoi{10.3847/1538-4357/ab6224}

\bibitem[{Kasper {et~al.}(2019)Kasper, Arsenault, Zins, Pantin, Duhoux,
  Riquelme, Kirchbauer, Pathak, Sterzik, Ageorges, Gutruf, Kampf, Maire, Huby,
  Guyon, Klupar, Mawet, Ruane, \& Karlsson}]{Kasper2019}
Kasper, M., Arsenault, R., Zins, G., {et~al.} 2019, The Messenger, 178, 5,
  \dodoi{10.18727/0722-6691/5163}

\bibitem[{Kervella {et~al.}(2017)Kervella, Th{\'{e}}venin, \&
  Lovis}]{Kervella2017}
Kervella, P., Th{\'{e}}venin, F., \& Lovis, C. 2017, Astron. Astrophys., 598,
  L7, \dodoi{10.1051/0004-6361/201629930}

\bibitem[{Koch \& Edvardsson(2002)}]{Koch2002}
Koch, A., \& Edvardsson, B. 2002, Astron. Astrophys., 381, 500,
  \dodoi{10.1051/0004-6361:20011578}

\bibitem[{Koll {et~al.}(2019)Koll, Malik, Mansfield, Kempton, Kite, Abbot, \&
  Bean}]{Koll2019}
Koll, D.~D., Malik, M., Mansfield, M., {et~al.} 2019, arXiv, 140,
  \dodoi{10.3847/1538-4357/ab4c91}

\bibitem[{Kopparapu {et~al.}(2013)Kopparapu, Ramirez, Kasting, Eymet, Robinson,
  Mahadevan, Terrien, Domagal-Goldman, Meadows, \& Deshpande}]{Kopparapu2013}
Kopparapu, R.~K., Ramirez, R., Kasting, J.~F., {et~al.} 2013, Astrophys. J.,
  765, \dodoi{10.1088/0004-637X/765/2/131}

\bibitem[{Kraus {et~al.}(2016)Kraus, Ireland, Huber, Mann, \&
  Dupuy}]{Kraus2016}
Kraus, A.~L., Ireland, M.~J., Huber, D., Mann, A.~W., \& Dupuy, T.~J. 2016,
  Astron. J., 152, 8, \dodoi{10.3847/0004-6256/152/1/8}

\bibitem[{Lasaga {et~al.}(1971)Lasaga, Holland, \& Dwyer}]{Lasaga1971}
Lasaga, A.~C., Holland, H.~D., \& Dwyer, M.~J. 1971, Science (80-. )., 174, 53,
  \dodoi{10.1126/science.174.4004.53}

\bibitem[{Li \& Fei(2014)}]{Li2014}
Li, J., \& Fei, Y. 2014, in Treatise on Geochemistry, 2nd edn., ed. R.~W.
  Carlson, Vol.~3 (Elsevier), 527--557,
  \dodoi{10.1016/B978-0-08-095975-7.00214-X}

\bibitem[{Lichtenberg {et~al.}(2019)Lichtenberg, Golabek, Burn, Meyer, Alibert,
  Gerya, \& Mordasini}]{Lichtenberg2019}
Lichtenberg, T., Golabek, G.~J., Burn, R., {et~al.} 2019, Nat. Astron., 3, 307,
  \dodoi{10.1038/s41550-018-0688-5}

\bibitem[{Liggins {et~al.}(2021)Liggins, Jordan, Rimmer, \&
  Shorttle}]{Liggins2021}
Liggins, P., Jordan, S., Rimmer, P.~B., \& Shorttle, O. 2021, 1.
\newblock \doarXiv{2111.05161}

\bibitem[{Lineweaver \& Chopra(2012)}]{Lineweaver2012}
Lineweaver, C.~H., \& Chopra, A. 2012, Annu. Rev. Earth Planet. Sci., 40, 597,
  \dodoi{10.1146/annurev-earth-042711-105531}

\bibitem[{Liu {et~al.}(2020)Liu, Yong, Asplund, Wang, Spina, Acu{\~{n}}a,
  Mel{\'{e}}ndez, \& Ram{\'{i}}rez}]{Liu2020}
Liu, F., Yong, D., Asplund, M., {et~al.} 2020, Mon. Not. R. Astron. Soc., 495,
  3961, \dodoi{10.1093/mnras/staa1420}

\bibitem[{Lodders(2003)}]{Lodders2003}
Lodders, K. 2003, Astrophys. J., 591, 1220, \dodoi{10.1086/375492}

\bibitem[{Lorenzo(2018)}]{Lorenzo2018}
Lorenzo, A.~J. 2018, Master's thesis, Arizona State University.
\newblock \url{https://repository.asu.edu/items/51736}

\bibitem[{Lu {et~al.}(2020)Lu, Schlaufman, \& Cheng}]{Lu2020}
Lu, C.~X., Schlaufman, K.~C., \& Cheng, S. 2020, Astron. J., 160, 253,
  \dodoi{10.3847/1538-3881/abb773}

\bibitem[{Luo {et~al.}(2016)Luo, Ono, Beukes, Wang, Xie, \& Summons}]{Luo2016}
Luo, G., Ono, S., Beukes, N.~J., {et~al.} 2016, Sci. Adv., 2, 1,
  \dodoi{10.1126/sciadv.1600134}

\bibitem[{Lupu {et~al.}(2014)Lupu, Zahnle, Marley, Schaefer, Fegley, Morley,
  Cahoy, Freedman, \& Fortney}]{Lupu2014}
Lupu, R.~E., Zahnle, K., Marley, M.~S., {et~al.} 2014, Astrophys. J., 784,
  \dodoi{10.1088/0004-637X/784/1/27}

\bibitem[{Lyons {et~al.}(2014)Lyons, Reinhard, \& Planavsky}]{Lyons2014}
Lyons, T.~W., Reinhard, C.~T., \& Planavsky, N.~J. 2014, Nature, 506, 307,
  \dodoi{10.1038/nature13068}

\bibitem[{Marty(2012)}]{Marty2012}
Marty, B. 2012, Earth Planet. Sci. Lett., 313-314, 56,
  \dodoi{10.1016/j.epsl.2011.10.040}

\bibitem[{McDonough(2003)}]{McDonough2003}
McDonough, W. 2003, in Treatise on Geochemistry, 1st edn., ed. R.~W. Carlson,
  Vol.~2 (Elsevier), 547--568, \dodoi{10.1016/B0-08-043751-6/02015-6}

\bibitem[{McDonough \& Sun(1995)}]{McDonough1995}
McDonough, W., \& Sun, S.-s. 1995, Chem. Geol., 120, 223,
  \dodoi{10.1016/0009-2541(94)00140-4}

\bibitem[{Moore {et~al.}(1998)Moore, Vennemann, \& Carmichael}]{Moore1998}
Moore, G., Vennemann, T., \& Carmichael, I. 1998, Am. Mineral., 83, 36

\bibitem[{Morel(2018)}]{Morel2018}
Morel, T. 2018, Astron. Astrophys., 615, A172,
  \dodoi{10.1051/0004-6361/201833125}

\bibitem[{Mulders {et~al.}(2016)Mulders, Pascucci, Apai, Frasca, \&
  Molenda-{\.{Z}}akowicz}]{Mulders2016}
Mulders, G.~D., Pascucci, I., Apai, D., Frasca, A., \& Molenda-{\.{Z}}akowicz,
  J. 2016, Astron. J., 152, 187, \dodoi{10.3847/0004-6256/152/6/187}

\bibitem[{Murakami {et~al.}(2002)Murakami, Hirose, Yurimoto, Nakashima, \&
  Takafuji}]{Murakami2002}
Murakami, M., Hirose, K., Yurimoto, H., Nakashima, S., \& Takafuji, N. 2002,
  Science (80-. )., 295, 1885, \dodoi{10.1126/science.1065998}

\bibitem[{{NASA Exoplanet Archive}(2022)}]{ps}
{NASA Exoplanet Archive}. 2022, Planetary Systems, Version: 2022-01-16 18:00,
  NExScI-Caltech/IPAC, \dodoi{10.26133/NEA12}

\bibitem[{Neuforge-Verheecke \& Magain(1997)}]{Neuforge1997}
Neuforge-Verheecke, C., \& Magain, P. 1997, Astron. Astrophys., 328, 261

\bibitem[{Nimmo {et~al.}(2020)Nimmo, Primack, Faber, Ramirez-Ruiz, \&
  Safarzadeh}]{Nimmo2020}
Nimmo, F., Primack, J., Faber, S.~M., Ramirez-Ruiz, E., \& Safarzadeh, M. 2020,
  Astrophys. J., 903, L37, \dodoi{10.3847/2041-8213/abc251}

\bibitem[{Noack {et~al.}(2012)Noack, Breuer, \& Spohn}]{Noack2012}
Noack, L., Breuer, D., \& Spohn, T. 2012, Icarus, 217, 484,
  \dodoi{10.1016/j.icarus.2011.08.026}

\bibitem[{Noack {et~al.}(2017)Noack, Rivoldini, \& {Van Hoolst}}]{Noack2017}
Noack, L., Rivoldini, A., \& {Van Hoolst}, T. 2017, Phys. Earth Planet. Inter.,
  269, 40, \dodoi{10.1016/j.pepi.2017.05.010}

\bibitem[{Norris \& Wood(2017)}]{Norris2017}
Norris, C.~A., \& Wood, B.~J. 2017, Nature, 549, 507,
  \dodoi{10.1038/nature23645}

\bibitem[{O'Brien {et~al.}(2018)O'Brien, Izidoro, Jacobson, Raymond, \&
  Rubie}]{OBrien2018}
O'Brien, D.~P., Izidoro, A., Jacobson, S.~A., Raymond, S.~N., \& Rubie, D.~C.
  2018, Space Sci. Rev., 214, \dodoi{10.1007/s11214-018-0475-8}

\bibitem[{O'Neill {et~al.}(2017)O'Neill, Marchi, Zhang, \& Bottke}]{ONeill2017}
O'Neill, C., Marchi, S., Zhang, S., \& Bottke, W. 2017, Nat. Geosci., 10, 793,
  \dodoi{10.1038/ngeo3029}

\bibitem[{O'Neill(1987)}]{Oneill1987}
O'Neill, H. 1987, Am. Mineral., 72, 67

\bibitem[{O'Neill {et~al.}(2006)O'Neill, Berry, McCammon, Jayasuriya, Campbell,
  \& Foran}]{ONeill2006}
O'Neill, H. S.~C., Berry, A.~J., McCammon, C.~C., {et~al.} 2006, Am. Mineral.,
  91, 404, \dodoi{10.2138/am.2005.1929}

\bibitem[{Ortenzi {et~al.}(2020)Ortenzi, Noack, Sohl, Guimond, Grenfell, Dorn,
  Schmidt, Vulpius, Katyal, Kitzmann, \& Rauer}]{Ortenzi2020}
Ortenzi, G., Noack, L., Sohl, F., {et~al.} 2020, Sci. Rep., 10, 1,
  \dodoi{10.1038/s41598-020-67751-7}

\bibitem[{Palme \& O'Neill(2014)}]{Palme2014b}
Palme, H., \& O'Neill, H. 2014, in Treatise on Geochemistry, 2nd edn., ed.
  R.~W. Carlson, Vol.~3 (Elsevier), 1--39,
  \dodoi{10.1016/B978-0-08-095975-7.00201-1}

\bibitem[{Pearson {et~al.}(2014)Pearson, Brenker, Nestola, McNeill, Nasdala,
  Hutchison, Matveev, Mather, Silversmit, Schmitz, Vekemans, \&
  Vincze}]{Pearson2014}
Pearson, D.~G., Brenker, F.~E., Nestola, F., {et~al.} 2014, Nature, 507, 221,
  \dodoi{10.1038/nature13080}

\bibitem[{Peslier {et~al.}(2017)Peslier, Sch{\"{o}}nb{\"{a}}chler, Busemann, \&
  Karato}]{Peslier2017}
Peslier, A.~H., Sch{\"{o}}nb{\"{a}}chler, M., Busemann, H., \& Karato, S.~I.
  2017, Space Sci. Rev., 212, 743, \dodoi{10.1007/s11214-017-0387-z}

\bibitem[{Petigura {et~al.}(2018)Petigura, Marcy, Winn, Weiss, Fulton, Howard,
  Sinukoff, Isaacson, Morton, \& Johnson}]{Petigura2018}
Petigura, E.~A., Marcy, G.~W., Winn, J.~N., {et~al.} 2018, Astron. J., 155, 89,
  \dodoi{10.3847/1538-3881/aaa54c}

\bibitem[{Phillips \& Russell(1987)}]{Phillips1987}
Phillips, J.~L., \& Russell, C.~T. 1987, Adv. Sp. Res., 7, 291,
  \dodoi{10.1016/0273-1177(87)90232-8}

\bibitem[{Pourbaix \& Boffin(2016)}]{Pourbaix2016}
Pourbaix, D., \& Boffin, H.~M. 2016, Astron. Astrophys., 586, 4,
  \dodoi{10.1051/0004-6361/201527859}

\bibitem[{Quanz {et~al.}(2015)Quanz, Crossfield, Meyer, Schmalzl, \&
  Held}]{Quanz2015}
Quanz, S.~P., Crossfield, I., Meyer, M.~R., Schmalzl, E., \& Held, J. 2015,
  Int. J. Astrobiol., 14, 279, \dodoi{10.1017/S1473550414000135}

\bibitem[{Quanz {et~al.}(2021)Quanz, Ottiger, Fontanet, Kammerer, Menti,
  Dannert, Gheorghe, Absil, Airapetian, Alei, Allart, Angerhausen, Blumenthal,
  Buchhave, Cabrera, Carri{\'{o}}n-Gonz{\'{a}}lez, Chauvin, Danchi, Dandumont,
  Defr{\`{e}}re, Dorn, Ehrenreich, Ertel, Fridlund, Mu{\~{n}}oz, Gasc{\'{o}}n,
  Girard, Glauser, Grenfell, Guidi, Hagelberg, Helled, Ireland, Kopparapu,
  Korth, Kozakis, Kraus, L{\'{e}}ger, Leedj{\"{a}}rv, Lichtenberg, Lillo-Box,
  Linz, Liseau, Loicq, Mahendra, Malbet, Mathew, Mennesson, Meyer, Mishra,
  Molaverdikhani, Noack, Oza, Pall{\'{e}}, Parviainen, Quirrenbach, Rauer,
  Ribas, Rice, Romagnolo, Rugheimer, Schwieterman, Serabyn, Sharma, Stassun,
  Szul{\'{a}}gyi, Wang, Wunderlich, Wyatt, \& Collaboration}]{Quanz2021}
Quanz, S.~P., Ottiger, M., Fontanet, E., {et~al.} 2021.
\newblock \doarXiv{2101.07500}

\bibitem[{Quarles {et~al.}(2020)Quarles, Li, Kostov, \&
  Haghighipour}]{Quarles2020}
Quarles, B., Li, G., Kostov, V., \& Haghighipour, N. 2020, Astron. J., 159, 80,
  \dodoi{10.3847/1538-3881/ab64fa}

\bibitem[{Quarles \& Lissauer(2016)}]{Quarles2016}
Quarles, B., \& Lissauer, J.~J. 2016, Astron. J., 151, 111,
  \dodoi{10.3847/0004-6256/151/5/111}

\bibitem[{Rajpaul {et~al.}(2016)Rajpaul, Aigrain, \& Roberts}]{Rajpaul2016}
Rajpaul, V., Aigrain, S., \& Roberts, S. 2016, Mon. Not. R. Astron. Soc. Lett.,
  456, L6, \dodoi{10.1093/mnrasl/slv164}

\bibitem[{Salmon {et~al.}(2021)Salmon, {Van Grootel}, Buldgen, Dupret, \&
  Eggenberger}]{Salmon2021}
Salmon, S.~J., {Van Grootel}, V., Buldgen, G., Dupret, M.~A., \& Eggenberger,
  P. 2021, Astron. Astrophys., 646, \dodoi{10.1051/0004-6361/201937174}

\bibitem[{Savel {et~al.}(2020)Savel, Dressing, Hirsch, Ciardi, Fleming,
  Giacalone, Mayo, \& Christiansen}]{Savel2020}
Savel, A.~B., Dressing, C.~D., Hirsch, L.~A., {et~al.} 2020, Astron. J., 160,
  287, \dodoi{10.3847/1538-3881/abc47d}

\bibitem[{Schubert \& Soderlund(2011)}]{Schubert2011}
Schubert, G., \& Soderlund, K.~M. 2011, Phys. Earth Planet. Inter., 187, 92,
  \dodoi{10.1016/j.pepi.2011.05.013}

\bibitem[{Schulze {et~al.}(2021)Schulze, Wang, Johnson, Gaudi, Unterborn, \&
  Panero}]{Schulze2021}
Schulze, J.~G., Wang, J., Johnson, J.~A., {et~al.} 2021, Planet. Sci. J., 2,
  113, \dodoi{10.3847/psj/abcaa8}

\bibitem[{Shah {et~al.}(2020)Shah, Alibert, Helled, \& Mezger}]{Shah2021}
Shah, O., Alibert, Y., Helled, R., \& Mezger, K. 2020, 162.
\newblock \doarXiv{2012.06455}

\bibitem[{Sossi {et~al.}(2020)Sossi, Burnham, Badro, Lanzirotti, Newville, \&
  Neill}]{Sossi2020}
Sossi, P.~A., Burnham, A.~D., Badro, J., {et~al.} 2020, 1

\bibitem[{Sossi \& Fegley(2018)}]{Sossi2018}
Sossi, P.~A., \& Fegley, B. 2018, Rev. Mineral. Geochemistry, 84, 393,
  \dodoi{10.2138/rmg.2018.84.11}

\bibitem[{Sossi {et~al.}(2019)Sossi, Klemme, O'Neill, Berndt, \&
  Moynier}]{Sossi2019}
Sossi, P.~A., Klemme, S., O'Neill, H. S.~C., Berndt, J., \& Moynier, F. 2019,
  Geochim. Cosmochim. Acta, 260, 204, \dodoi{10.1016/j.gca.2019.06.021}

\bibitem[{Spaargaren {et~al.}(2020)Spaargaren, Wang, Ballmer, Mojzsis, \&
  Tackley}]{Spaargaren2020}
Spaargaren, R., Wang, H., Ballmer, M., Mojzsis, S., \& Tackley, P. 2020, in
  Eur. Sci. Congr. 2020 (https://doi.org/10.5194/epsc2020-745),
  \dodoi{10.5194/epsc2020-745}

\bibitem[{Stixrude \& Lithgow-Bertelloni(2011)}]{Stixrude2011}
Stixrude, L., \& Lithgow-Bertelloni, C. 2011, Geophys. J. Int., 184, 1180,
  \dodoi{10.1111/j.1365-246X.2010.04890.x}

\bibitem[{Su{\'{a}}rez-Andr{\'{e}}s {et~al.}(2018)Su{\'{a}}rez-Andr{\'{e}}s,
  Israelian, {Gonz{\'{a}}lez Hern{\'{a}}ndez}, Adibekyan, {Delgado Mena},
  Santos, \& Sousa}]{Suarez-Andres2018}
Su{\'{a}}rez-Andr{\'{e}}s, L., Israelian, G., {Gonz{\'{a}}lez Hern{\'{a}}ndez},
  J.~I., {et~al.} 2018, Astron. Astrophys., 614, A84,
  \dodoi{10.1051/0004-6361/201730743}

\bibitem[{{The LUVIOR team}(2019)}]{LUVOIR2019}
{The LUVIOR team}. 2019, {LUVOIR Final Report}, Tech. rep.
\newblock \url{https://arxiv.org/abs/1912.06219}

\bibitem[{Thiabaud {et~al.}(2014)Thiabaud, Marboeuf, Alibert, Cabral, Leya, \&
  Mezger}]{Thiabaud2014}
Thiabaud, A., Marboeuf, U., Alibert, Y., {et~al.} 2014, Astron. Astrophys.,
  562, A27, \dodoi{10.1051/0004-6361/201322208}

\bibitem[{Thiabaud {et~al.}(2015)Thiabaud, Marboeuf, Alibert, Leya, \&
  Mezger}]{Thiabaud2015}
Thiabaud, A., Marboeuf, U., Alibert, Y., Leya, I., \& Mezger, K. 2015, Astron.
  Astrophys., 580, A30, \dodoi{10.1051/0004-6361/201525963}

\bibitem[{Unterborn {et~al.}(2018)Unterborn, Desch, Hinkel, \&
  Lorenzo}]{Unterborn2018}
Unterborn, C.~T., Desch, S.~J., Hinkel, N.~R., \& Lorenzo, A. 2018, Nat.
  Astron., 2, 297, \dodoi{10.1038/s41550-018-0411-6}

\bibitem[{Unterborn {et~al.}(2016)Unterborn, Dismukes, \&
  Panero}]{Unterborn2016}
Unterborn, C.~T., Dismukes, E.~E., \& Panero, W.~R. 2016, Astrophys. J., 819,
  32, \dodoi{10.3847/0004-637x/819/1/32}

\bibitem[{Unterborn {et~al.}(2014)Unterborn, Kabbes, Pigott, Reaman, \&
  Panero}]{Unterborn2014}
Unterborn, C.~T., Kabbes, J.~E., Pigott, J.~S., Reaman, D.~M., \& Panero, W.~R.
  2014, Astrophys. J., 793, 124, \dodoi{10.1088/0004-637X/793/2/124}

\bibitem[{Wade \& Wood(2005)}]{Wade2005}
Wade, J., \& Wood, B.~J. 2005, Earth Planet. Sci. Lett., 236, 78,
  \dodoi{10.1016/j.epsl.2005.05.017}

\bibitem[{Wagner {et~al.}(2021)Wagner, Boehle, Pathak, Kasper, Arsenault,
  Jakob, K{\"{a}}ufl, Leveratto, Maire, Pantin, Siebenmorgen, Zins, Absil,
  Ageorges, Apai, Carlotti, Choquet, Delacroix, Dohlen, Duhoux, Forsberg,
  Fuenteseca, Gutruf, Guyon, Huby, Kampf, Karlsson, Kervella, Kirchbauer,
  Klupar, Kolb, Mawet, N'Diaye, {Orban de Xivry}, Quanz, Reutlinger, Ruane,
  Riquelme, Soenke, Sterzik, Vigan, \& de~Zeeuw}]{Wagner2021}
Wagner, K., Boehle, A., Pathak, P., {et~al.} 2021, Nat. Commun., 12, 1,
  \dodoi{10.1038/s41467-021-21176-6}

\bibitem[{Wang(2018)}]{Wang2018b}
Wang, H. 2018, Phd thesis, Australian National University,
  \dodoi{10.25911/5d5147caa0d1e}

\bibitem[{Wang {et~al.}(2020{\natexlab{a}})Wang, {T. Morel}, Quanz, \&
  Mojzsis}]{Wang2020}
Wang, H., {T. Morel}, Quanz, S., \& Mojzsis, S. 2020{\natexlab{a}}, Astron.
  Astrophys., 19, 1, \dodoi{10.1051/0004-6361/202038386}

\bibitem[{Wang {et~al.}(2018)Wang, Lineweaver, \& Ireland}]{Wang2018}
Wang, H.~S., Lineweaver, C.~H., \& Ireland, T.~R. 2018, Icarus, 299, 460,
  \dodoi{10.1016/j.icarus.2017.08.024}

\bibitem[{Wang {et~al.}(2019{\natexlab{a}})Wang, Lineweaver, \&
  Ireland}]{Wang2019a}
---. 2019{\natexlab{a}}, Icarus, 328, 287, \dodoi{10.1016/j.icarus.2019.03.018}

\bibitem[{Wang {et~al.}(2019{\natexlab{b}})Wang, Liu, Ireland, Brasser, Yong,
  \& Lineweaver}]{Wang2019b}
Wang, H.~S., Liu, F., Ireland, T.~R., {et~al.} 2019{\natexlab{b}}, Mon. Not. R.
  Astron. Soc., 482, 2222, \dodoi{10.1093/mnras/sty2749}

\bibitem[{Wang {et~al.}(2020{\natexlab{b}})Wang, Sossi, \& Quanz}]{Wang2020a}
Wang, H.~S., Sossi, P.~A., \& Quanz, S.~P. 2020{\natexlab{b}}, in Eur. Sci.
  Congr. 2020 (online: Europlanet Science Congress 2020), EPSC2020--874,
  \dodoi{10.5194/epsc2020-874}

\bibitem[{Warren \& Hauri(2014)}]{Warren2014}
Warren, J.~M., \& Hauri, E.~H. 2014, J. Geophys. Res. Solid Earth, 119, 1851,
  \dodoi{10.1002/2013JB010328}

\bibitem[{Winn \& Fabrycky(2015)}]{Winn2015}
Winn, J.~N., \& Fabrycky, D.~C. 2015, Annu. Rev. Astron. Astrophys., 53, 409,
  \dodoi{10.1146/annurev-astro-082214-122246}

\bibitem[{Wood {et~al.}(2019)Wood, Smythe, \& Harrison}]{Wood2019}
Wood, B.~J., Smythe, D.~J., \& Harrison, T. 2019, Am. Mineral., 104, 844,
  \dodoi{10.2138/am-2019-6852CCBY}

\bibitem[{Woodland \& Koch(2003)}]{Woodland2003}
Woodland, A.~B., \& Koch, M. 2003, Earth Planet. Sci. Lett., 214, 295,
  \dodoi{10.1016/S0012-821X(03)00379-0}

\bibitem[{Wu {et~al.}(2018)Wu, Desch, Schaefer, Elkins-Tanton, Pahlevan, \&
  Buseck}]{Wu2018}
Wu, J., Desch, S.~J., Schaefer, L., {et~al.} 2018, J. Geophys. Res. Planets,
  123, 2691, \dodoi{10.1029/2018JE005698}

\bibitem[{Yoshizaki \& McDonough(2020)}]{Yoshizaki2020}
Yoshizaki, T., \& McDonough, W.~F. 2020, Geochim. Cosmochim. Acta, 273, 137,
  \dodoi{10.1016/j.gca.2020.01.011}

\bibitem[{Zhao {et~al.}(2018)Zhao, Fischer, Brewer, Giguere, \&
  Rojas-Ayala}]{Zhao2018}
Zhao, L., Fischer, D.~A., Brewer, J., Giguere, M., \& Rojas-Ayala, B. 2018,
  Astron. J., 155, 24, \dodoi{10.3847/1538-3881/aa9bea}

\bibitem[{Zhu \& Dong(2021)}]{Zhu2021}
Zhu, W., \& Dong, S. 2021, Annu. Rev. Astron. Astrophys., 59, 291,
  \dodoi{10.1146/annurev-astro-112420-020055}

\bibitem[{Zuluaga \& Cuartas(2012)}]{Zuluaga2012}
Zuluaga, J.~I., \& Cuartas, P.~A. 2012, Icarus, 217, 88,
  \dodoi{10.1016/j.icarus.2011.10.014}

\end{thebibliography}

\end{document}